\documentclass[11pt, fleqn, a4paper]{article}
\usepackage[usenames]{color} 
\usepackage{amssymb} 
\usepackage{amsmath} 
\usepackage[utf8]{inputenc} 
\usepackage{braket, tensor, bm}
\usepackage[dvipdfmx]{graphicx}
\usepackage{mediabb}
\usepackage{here}
\usepackage{mathptmx}
\usepackage{algorithm,algorithmic}  
\usepackage{here}
\usepackage{authblk}
\usepackage{cite}

\makeatletter
\@addtoreset{equation}{section}

\@addtoreset{figure}{section}

\@addtoreset{table}{section}
\makeatother

\title{A Parallel Computing Method for the Higher Order Tensor Renormalization Group}
\author[1]{Takumi Yamashita}  \author[2]{Tetsuya Sakurai}
\affil[1]{{\small Center for Artificial Intelligence Research, University of Tsukuba, 1-1-1 Tennodai, Tsukuba, Ibaraki 305-8577 Japan, e-mail: yamasita@ccs.tsukuba.ac.jp}}
\affil[2]{{\small Department of Computer Science, University of Tsukuba, 1-1-1 Tennodai, Tsukuba, Ibaraki 305-8573 Japan, e-mail: sakurai@cs.tsukuba.ac.jp}}
\date{}

\begin{document}
 
 \parindent = 10pt
  
  \maketitle
  
  \begin{abstract} 
    In this paper, we propose a parallel computing method for the Higher Order Tensor Renormalization Group (HOTRG) applied to a $d$-dimensional $( d \geq 2 )$ simple lattice model.
    Sequential computation of the HOTRG requires $O ( \chi^{4 d - 1} )$ computational cost, where $\chi$ is bond dimension, in a step to contract indices of tensors.
    When we simply distribute elements of a local tensor to each process in parallel computing of the HOTRG, frequent communication between processes occurs.
    The simplest way to avoid such communication is to hold all the tensor elements in each process, however, it requires $O ( \chi^{2d} )$ memory space.
    In the presented method, placement of a local tensor element to more than one process is accepted and sufficient local tensor elements are distributed to each process to avoid communication between processes during considering computation step.
    For the bottleneck part of computational cost,
    such distribution is achieved by distributing elements of  two local tensors to $\chi^2$ processes according to one of the indices of each local tensor which are not contracted during considering computation.
    In the case of $d \geq 3$, computational cost in each process is reduced to $O ( \chi^{4 d - 3} )$ and memory space requirement in each process is kept to be $O ( \chi^{2d - 1} )$.
  \end{abstract}

 \section{Introduction}  \label{Introduction}

  Thermodynamic properties in a lattice model have been studied vigorously.
  The tensor renormalization group (TRG) method \cite{LN07} is a powerful technique for such study.
  In this method, the partition function is formulated by using tensor network.
  Exact computation of the partition function from a tensor network requires vast computational cost.
  As an alternative, the TRG approximates the partition function through a procedure called coarse-graining.
  In this procedure, a tensor network is updated as a coarser tensor network through singular value decomposition (SVD).
  Xie {\it et al}. \cite{XCQZYX12} proposed another improved TRG method using higher-order singular value decomposition (HOSVD) \cite{LMV00}
  for approximation of the partition function.
  In \cite{XCQZYX12}, it is abbreviated as HOTRG.
  The HOTRG is applicable not only to a two-dimensional lattice model but also to higher-dimensional one.
  The HOTRG has been applied to several kinds of physics models \cite{CLXHHCWXX17, CXY18, GGN16, KGLCTG18, KT16, QCCXKZBX13, WXCNX14, YXMLDZQCX14}.
  Details of the HOTRG are not explicitly shown in \cite{XCQZYX12}, however, it is readily deduced.
  A way of implementation of this method is not unique.
  A comclete example is shown in \cite{YIIS18}.
  On one hand, the HOTRG has the above-mentioned merit, but on the other hand, computational cost and memory space requirement of it in higher-dimensional simple lattice model are far from cheap.
  In a $d$-dimensional simple lattice model, computational cost and memory space requirement are $O ( \chi^{4d - 1} )$ and $O ( \chi^{2d} )$, respectively, where $\chi$ is bond dimension of indices of a local tensor.
   When we consider parallel computing of the HOTRG, another problem occurs if we simply distribute local tensor elements to each process.
   In the simplest way of distribution, one local tensor element is placed to one process.
   In such a case, necessary local tensor elements in contraction procedures are placed more than one process and cost for communication between processes in contraction is a problem.

  In this paper, we propose a parallel computing method for the HOTRG which avoid the problem of the cost for communication.
  We can avoid the problem if sufficient local tensor elements for a considering contraction procedure are placed to one process.
  Then, we have no reason to persist in a rule that an element of a local tensor element is placed to one process.
  In other words, we accept that an element of a local tensor is placed to more than one process.
  In development of our method, contraction procedures which requires elements of two local tensors are considered.
  During considering contraction procedure, we focus on one of indices of each local tensor.
  These indices have a characteristic that they are not contracted.
  Let us denote the focused indices of local tensor $T_1$ and $T_2$ by $i_1$ and $i_2$, respectively.
  For a specified bond dimension $\chi$, our method use $\chi^2$ processes and let their process numbers be expressed as $( p_1 + p_2 \chi )$ $( p_1, p_2 = 0, 1, ..., \chi - 1 )$.
  Then, elements of local tensor $T_1$ whose index $i_1$ is $i_1 = \nu$ are placed to processes whose process number satisfies $p_1 = \nu$.
  Placement of elements of local tensor $T_2$ is similar.
  Thus, sufficient local tensor elements for contraction are placed to one process and we can avoid the problem of cost for communication between processes.
  As further advantages, in the cases of $d \geq 3$, computational cost in each process is reduced from $O ( \chi^{4d - 1} )$ to $O ( \chi^{4d - 3} )$
  since the bottleneck part in computational cost is executed in parallel in $\chi^2$ processes and memory space requirement in each process is reduced from $O ( \chi^{2d} )$ to $O ( \chi^{2d - 1} )$.
  More importantly, key ideas in our method, distribution of sufficient tensor elements to each process and a way of distribution of tensor elements according to indices which are not contracted during considering contraction step,
  can be applicable to another method if it has mathematical structure which is suitable for these ideas.

   This paper is organized as follows.
   In Section \ref{HOTRG}, the HOTRG is described.
   In Section \ref{ParallelMethod}, we explain key ideas of our parallel computing method for the HOTRG.
   In Section \ref{Implementation}, we give a way to implement our method.
   In Section \ref{NumExp}, numerical experiments are given.
   Section \ref{ConcludingRmks} is concluding remarks.

\section{The HOTRG method}  \label{HOTRG}

  The HOTRG method is presented by Xie {\it et al}. \cite{XCQZYX12}.
  Here we introduce it briefly.

  Let us consider a model in $d$-dimensional simple lattice with periodic boundary condition such that its Hamiltonian $H$ is given in the form of
  \begin{equation}
      - \frac{H}{k_B T} = \sum_{\langle i, j \rangle} f ( \sigma_i, \sigma_j ) + \sum_{i = 1}^N g ( \sigma_i ),
  \end{equation}
  where $\sigma_i$ is a spin degree of freedom in site $i$, $k_B$ is the Boltzmann constant and $T$ is temperature.
  Then, partition function $Z$ of this model is given as
  \begin{equation}
      Z = \sum_{\sigma_i} \prod_{\langle i, j \rangle} e^{f ( \sigma_i, \sigma_j )} \prod_{i = 1}^N e^{g ( \sigma_i )}
         = \sum_{\sigma_i} \prod_{\langle i, j \rangle} W_{\sigma_i \sigma_j} \prod_{i = 1}^N V_{\sigma_i}.
  \end{equation}
  Representation of partition function using tensor network is given as
  \begin{equation}
      Z = \textrm{tTr} \prod_{i = 1}^N T_{j_1 j_1^{\prime} ... j_d j_d^{\prime}}^i,
  \end{equation}
  where $T_{j_1 j_1^{\prime} ... j_d j_d^{\prime} }^i$ is a local tensor in site $i$ and $\textrm{tTr}$ means that we total over all the combinations of indices.
  See also \cite{ZXCWCX10}.
  From eigenvalue decomposition, we have $W = U \Lambda U^{\dagger}$.
  Introducing a matrix $X$ given as $X = U \sqrt{\Lambda}$, we have
  \begin{equation}
      T_{j_1 j_1^{\prime} ... j_d j_d^{\prime}}^i = \sum_{s = 0}^{q - 1} X_{s j_1} \cdots X_{s j_d} X_{s j_1^{\prime}}^* \cdots X_{s j_d^{\prime}}^* V_s,
  \end{equation}
  where $q$ is number of states which a spin degree of freedom takes.

\begin{figure}[H]
      \begin{center}
          \includegraphics[width=0.6\textwidth]{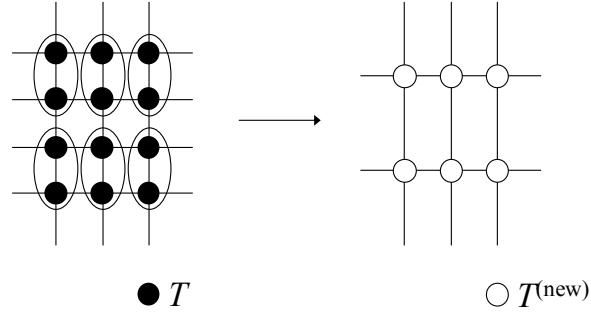}  \\
          \caption{Description of coarse-graining as tensor network}
           \label{Fig_CoarseGraining}
      \end{center}
  \end{figure}

  In the HOTRG, a procedure called coarse-graining is repeatedly applied to a tensor network to compute partition function $Z$ approximately.
  This procedure is to merge two neighboring local tensors approximately into one new local tensor.
  The number of considering sites is reduced by half.
  Representation of this procedure as a tensor network is given in Fig. \ref{Fig_CoarseGraining}.
  Figures given in this section describe the case of two-dimensional lattice for simplicity.
  Expansion to higher-dimensional lattice is straightforward.
  Coarse-graining procedure is applied to each direction of a lattice by turns.

\begin{figure}[H]
      \begin{center}
          \includegraphics[width=0.25\textwidth]{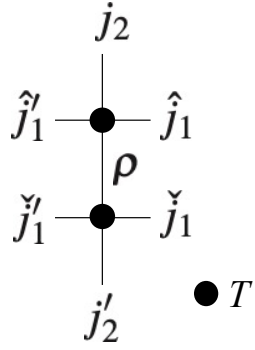}  \\
          \caption{Description of tensor $M$ as tensor network}
           \label{Fig_TensorM}
      \end{center}
  \end{figure}

\begin{figure}[H]
      \begin{center}
          \includegraphics[width=0.4\textwidth]{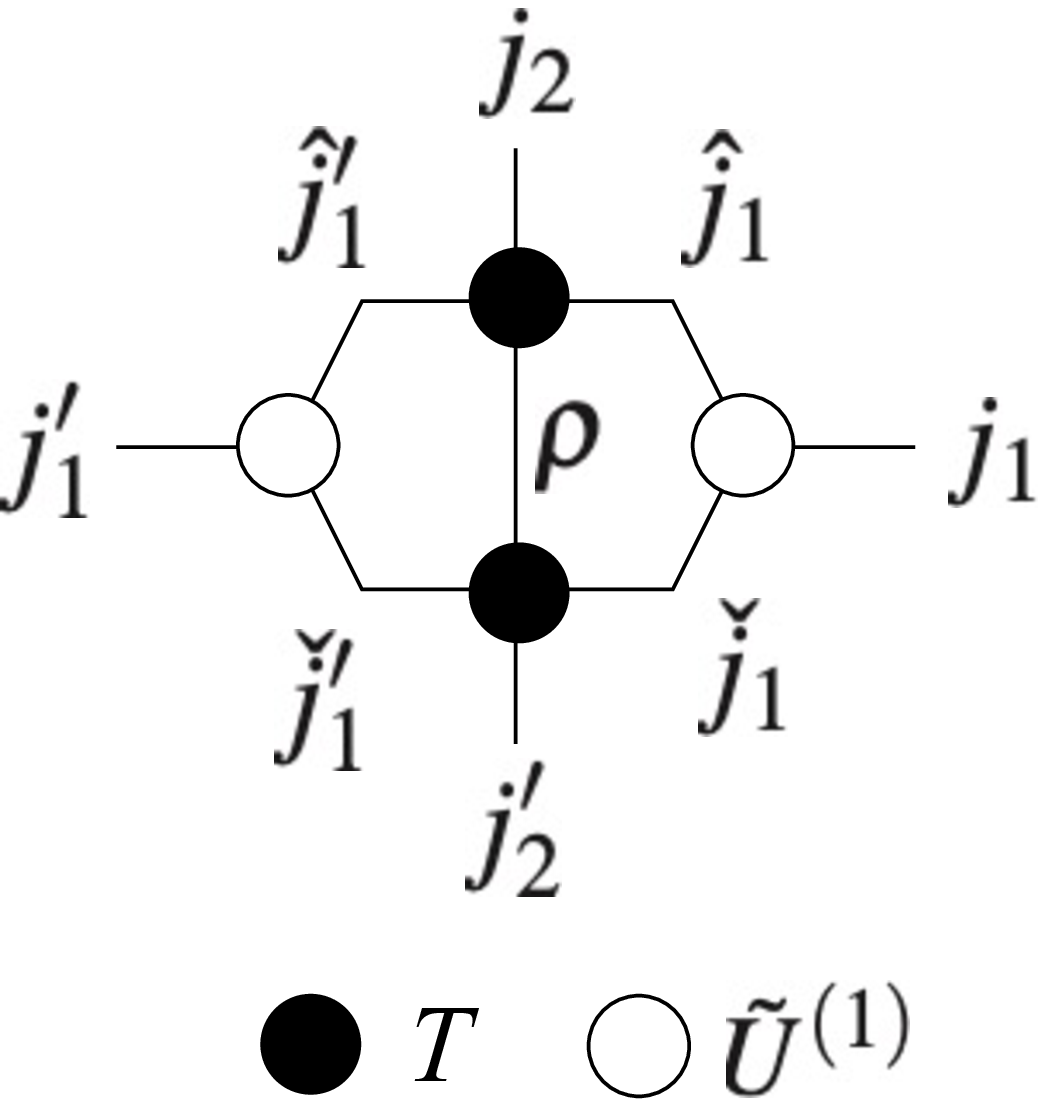}  \\
          \caption{Description of coarse-graining procedure (\ref{CoarseGrainingProcedure_d}) as tensor network}
           \label{Fig_NewTensor}
      \end{center}
  \end{figure}

  For local tensors $T_{j_1 j_1^{\prime} ... j_d j_d^{\prime}}$, assume that the indices $j_s$ and $j_s^{\prime}$ take values $0, 1, ..., \chi_s - 1$ $( s = 1, ..., d )$.
  Moreover, assume that coarse-graining procedure is applied to the direction represented by indices $j_d$ and $j_d^{\prime}$ and new local tensors $T_{j_1 j_1^{\prime} ... j_d j_d^{\prime}}^{( \textrm{new} )}$ are constructed.
  This procedure is mathematically expressed as
  \begin{align}
      T_{j_1 j_1^{\prime} ... j_d j_d^{\prime}}^{( \textrm{new} )}
      = \sum &\tilde{U}_{( \hat{j}_1 \check{j}_1 ), j_1}^{( 1 )} \cdots \tilde{U}_{( \hat{j}_{d - 1} \check{j}_{d - 1} ), j_{d - 1}}^{( d - 1 )}  \notag  \\
                  &\times M_{( \hat{j}_1 \check{j}_1 ) ( \hat{j}_1^{\prime} \check{j}_1^{\prime} ) ... ( \hat{j}_{d - 1} \check{j}_{d - 1} ) ( \hat{j}_{d - 1}^{\prime} \check{j}_{d - 1}^{\prime} ) j_d j_d^{\prime}}  \notag  \\     
                  &\times \tilde{U}_{( \hat{j}_1^{\prime} \check{j}_1^{\prime} ), j_1^{\prime}}^{( 1 )} \cdots \tilde{U}_{( \hat{j}_{d - 1}^{\prime} \check{j}_{d - 1}^{\prime} ), j_{d - 1}^{\prime}}^{( d - 1 )},  \label{CoarseGrainingProcedure_d}
  \end{align}
  where $\sum$ is
  \begin{equation}
      \sum_{\hat{j}_1 = 0}^{\chi_1 - 1} \sum_{\check{j}_1 = 0}^{\chi_1 - 1} \cdots \sum_{\hat{j}_{d - 1} = 0}^{\chi_{d - 1} - 1}  \sum_{\check{j}_{d - 1} = 0}^{\chi_{d - 1} - 1}
      \sum_{\hat{j}_1^{\prime} = 0}^{\chi_1 - 1} \sum_{\check{j}_1^{\prime} = 0}^{\chi_1 - 1} \cdots \sum_{\hat{j}_{d - 1}^{\prime} = 0}^{\chi_{d - 1} - 1}  \sum_{\check{j}_{d - 1}^{\prime} = 0}^{\chi_{d - 1} - 1},
  \end{equation}
  tensor $M_{( \hat{j}_1 \check{j}_1 ) ( \hat{j}_1^{\prime} \check{j}_1^{\prime} ) ... ( \hat{j}_{d - 1} \check{j}_{d - 1} ) ( \hat{j}_{d - 1}^{\prime} \check{j}_{d - 1}^{\prime} ) j_d j_d^{\prime}} $ is
  \begin{equation}
      M_{( \hat{j}_1 \check{j}_1 ) ( \hat{j}_1^{\prime} \check{j}_1^{\prime} ) ... ( \hat{j}_{d - 1} \check{j}_{d - 1} ) ( \hat{j}_{d - 1}^{\prime} \check{j}_{d - 1}^{\prime} ) j_d j_d^{\prime}} 
      = \sum_{\rho = 0}^{\chi_d - 1} T_{\hat{j}_1 \hat{j}_1^{\prime} ... \hat{j}_{d - 1} \hat{j}_{d - 1}^{\prime} j_d \rho} T_{\check{j}_1 \check{j}_1^{\prime} ... \check{j}_{d - 1} \check{j}_{d - 1}^{\prime} \rho j_d^{\prime}},
  \end{equation}
  and $\tilde{U}_{( \hat{j}_s \check{j}_s ), j_s}^{( s )}$ and $\tilde{U}_{( \hat{j}_s^{\prime} \check{j}_s^{\prime} ), j_s^{\prime}}^{( s )}$ $( s = 1, ..., d - 1 )$ are unitary matrices.
  For a parameter $\chi$ which we specify,
  the indices $j_s$ and $j_s^{\prime}$ $( s = 1, ..., d - 1 )$ of tensor $T_{j_1 j_1^{\prime} ... j_d j_d^{\prime}}^{( \textrm{new} )}$ take values $0, 1, ..., \chi_s^{( \textrm{new} )} - 1$, where $\chi_s^{( \textrm{new} )}$ is $\min \{ \chi_s^2, \chi \}$.
  It means that unitary matrices $\tilde{U}_{( \hat{j}_s \check{j}_s ), j_s}^{( s )}$ and $\tilde{U}_{( \hat{j}_s^{\prime} \check{j}_s^{\prime} ), j_s^{\prime}}^{( s )}$ are truncated and only the first $\chi$ columns are considered if it holds $\chi_s^2 > \chi$.
  Accuracy of approximated partition function depends on the parameter $\chi$.
  Representations of tensor $M_{( \hat{j}_1 \check{j}_1 ) ( \hat{j}_1^{\prime} \check{j}_1^{\prime} ) ... ( \hat{j}_{d - 1} \check{j}_{d - 1} ) ( \hat{j}_{d - 1}^{\prime} \check{j}_{d - 1}^{\prime} ) j_d j_d^{\prime}} $
                                 and coarse-graining procedure (\ref{CoarseGrainingProcedure_d}) as tensor network are shown in Figs. \ref{Fig_TensorM} and \ref{Fig_NewTensor}, respectively.
  The indices $j_d$ and $j_d^{\prime}$ of tensor $T_{j_1 j_1^{\prime} ... j_d j_d^{\prime}}^{( \textrm{new} )}$ take values $0, 1, ..., \chi_d - 1$.

\begin{figure}[H]
      \begin{center}
          \includegraphics[width=0.6\textwidth]{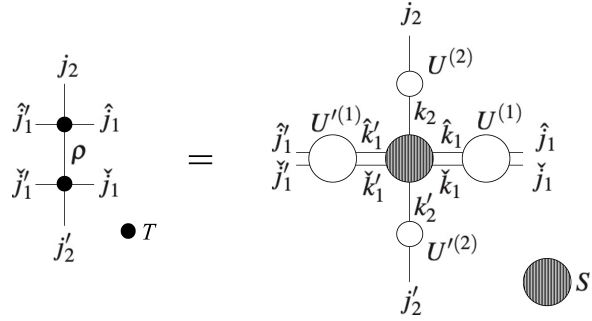}  \\
          \caption{Description of the HOSVD of tensor $M$ as tensor network}
           \label{Fig_HOSVD}
      \end{center}
  \end{figure}

  Next, we explain a method to obtain unitary matrices $\tilde{U}_{( \hat{j}_s \check{j}_s ), j_s}^{( s )}$ and $\tilde{U}_{( \hat{j}_s^{\prime} \check{j}_s^{\prime} ), j_s^{\prime}}^{( s )}$ $( s = 1, ..., d - 1 )$.
  Tensor $M_{( \hat{j}_1 \check{j}_1 ) ( \hat{j}_1^{\prime} \check{j}_1^{\prime} ) ... ( \hat{j}_{d - 1} \check{j}_{d - 1} ) ( \hat{j}_{d - 1}^{\prime} \check{j}_{d - 1}^{\prime} ) j_d j_d^{\prime}}$
  is decomposed by Higher Order Singular Value Decomposition (HOSVD) \cite{LMV00} as
  \begin{align}
      &M_{( \hat{j}_1 \check{j}_1 ) ( \hat{j}_1^{\prime} \check{j}_1^{\prime} ) ... ( \hat{j}_{d - 1} \check{j}_{d - 1} ) ( \hat{j}_{d - 1}^{\prime} \check{j}_{d - 1}^{\prime} ) j_d j_d^{\prime}}  \notag  \\
      &= \sum S_{( \hat{k}_1 \check{k}_1 ) ( \hat{k}_1^{\prime} \check{k}_1^{\prime} ) ... ( \hat{k}_{d - 1} \check{k}_{d - 1} ) ( \hat{k}_{d - 1}^{\prime} \check{k}_{d - 1}^{\prime} ) k_d k_d^{\prime}}  \notag  \\
      &  ~~~~~~~~~
         \times  U_{( \hat{j}_1 \check{j}_1 ), ( \hat{k}_1 \check{k}_1 )}^{( 1 )} \cdots U_{( \hat{j}_{d - 1} \check{j}_{d - 1} ), ( \hat{k}_{d - 1} \check{k}_{d - 1} )}^{( d - 1 )} U_{j_d, k_d}^{( d )}  \notag  \\
      &  ~~~~~~~~~
         \times  U_{( \hat{j}_1^{\prime} \check{j}_1^{\prime} ), ( \hat{k}_1^{\prime} \check{k}_1^{\prime} )}^{\prime ( 1 )}
                     \cdots
                     U_{( \hat{j}_{d - 1}^{\prime} \check{j}_{d - 1}^{\prime} ), ( \hat{k}_{d - 1}^{\prime} \check{k}_{d - 1}^{\prime} )}^{\prime ( d - 1 )} U_{j_d^{\prime}, k_d^{\prime}}^{\prime ( d )},  \label{Eq_HOSVD}
  \end{align}
  where $\sum$ is
  \begin{equation}
      \sum = \sum_{\hat{k}_1 = 0}^{\chi_1 - 1} \sum_{\check{k}_1 = 0}^{\chi_1 - 1} \cdots \sum_{\hat{k}_{d - 1} = 0}^{\chi_{d - 1} - 1} \sum_{\check{k}_{d - 1} = 0}^{\chi_{d - 1} - 1}
                  \sum_{\hat{k}_1^{\prime} = 0}^{\chi_1 - 1} \sum_{\check{k}_1^{\prime} = 0}^{\chi_1 - 1} \cdots \sum_{\hat{k}_{d - 1}^{\prime} = 0}^{\chi_{d - 1} - 1} \sum_{\check{k}_{d - 1}^{\prime} = 0}^{\chi_{d - 1} - 1}
                  \sum_{k_d = 0}^{\chi_d - 1} \sum_{k_d^{\prime} = 0}^{\chi_d - 1},
  \end{equation}
  $S_{( \hat{k}_1 \check{k}_1 ) ( \hat{k}_1^{\prime} \check{k}_1^{\prime} ) ... ( \hat{k}_{d - 1} \check{k}_{d - 1} ) ( \hat{k}_{d - 1}^{\prime} \check{k}_{d - 1}^{\prime} ) k_d k_d^{\prime}}$
  is the core tensor of $M_{( \hat{j}_1 \check{j}_1 ) ( \hat{j}_1^{\prime} \check{j}_1^{\prime} ) ... ( \hat{j}_{d - 1} \check{j}_{d - 1} ) ( \hat{j}_{d - 1}^{\prime} \check{j}_{d - 1}^{\prime} ) j_d j_d^{\prime}} $,
  and $U_{( \hat{j}_s \check{j}_s ), ( \hat{k}_s \check{k}_s )}^{( s )}$,
         $U_{( \hat{j}_s^{\prime} \check{j}_s^{\prime} ), ( \hat{k}_s^{\prime} \check{k}_s^{\prime} )}^{\prime ( s )}$ $( s = 1, ..., d - 1 )$,
         $U_{j_d, k_d}^{( d )}$ and $U_{j_d^{\prime}, k_d^{\prime}}^{\prime ( d )}$ are unitary matrices.
  For $s = 1, ..., d - 1$, one of $U_{( \hat{j}_s \check{j}_s ), ( \hat{k}_s \check{k}_s )}^{( s )}$ and $U_{( \hat{j}_s^{\prime} \check{j}_s^{\prime} ), ( \hat{k}_s^{\prime} \check{k}_s^{\prime} )}^{\prime ( s )}$
                                     is chosen according to some standard and used as $\tilde{U}_{( \hat{j}_s \check{j}_s ), j_s}^{( s )}$ and $\tilde{U}_{( \hat{j}_s^{\prime} \check{j}_s^{\prime} ), j_s^{\prime}}^{( s )}$ in (\ref{CoarseGrainingProcedure_d}).
  Representation of decomposition (\ref{Eq_HOSVD}) as tensor network is shown in Fig. \ref{Fig_HOSVD}.

  \noindent
  We can obtain unitary matrices $U_{( \hat{j}_s \check{j}_s ), ( \hat{k}_s \check{k}_s )}^{( s )}$ and $U_{( \hat{j}_s^{\prime} \check{j}_s^{\prime} ), ( \hat{k}_s^{\prime} \check{k}_s^{\prime} )}^{\prime ( s )}$ $( s = 1, ..., d - 1 )$ without execution of the HOSVD.
  We explain the case of $U_{( \hat{j}_1 \check{j}_1 ), ( \hat{k}_1 \check{k}_1 )}^{( 1 )}$ and $U_{( \hat{j}_1^{\prime} \check{j}_1^{\prime} ), ( \hat{k}_1^{\prime} \check{k}_1^{\prime} )}^{\prime ( 1 )}$.
  The other cases are similar.
  Let us introduce matrices
  \begin{equation}
      \tilde{M}_{( \hat{j}_1 \check{j}_1 ), ( \hat{j}_1^{\prime} \check{j}_1^{\prime} ) ( \hat{j}_2 \check{j}_2 ) ( \hat{j}_2^{\prime} \check{j}_2^{\prime} ) ... ( \hat{j}_{d - 1} \check{j}_{d - 1} ) ( \hat{j}_{d - 1}^{\prime} \check{j}_{d - 1}^{\prime} ) j_d j_d^{\prime}}^{( 1 )}
      = M_{( \hat{j}_1 \check{j}_1 ) ( \hat{j}_1^{\prime} \check{j}_1^{\prime} ) ... ( \hat{j}_{d - 1} \check{j}_{d - 1} ) ( \hat{j}_{d - 1}^{\prime} \check{j}_{d - 1}^{\prime} ) j_d j_d^{\prime}}
  \end{equation}
  and
  \begin{equation}
      \tilde{M}_{( \hat{j}_1^{\prime} \check{j}_1^{\prime} ), ( \hat{j}_1 \check{j}_1 ) ( \hat{j}_2 \check{j}_2 ) ( \hat{j}_2^{\prime} \check{j}_2^{\prime} ) ... ( \hat{j}_{d - 1} \check{j}_{d - 1} ) ( \hat{j}_{d - 1}^{\prime} \check{j}_{d - 1}^{\prime} ) j_d j_d^{\prime}}^{\prime ( 1 )}
      = M_{( \hat{j}_1 \check{j}_1 ) ( \hat{j}_1^{\prime} \check{j}_1^{\prime} ) ... ( \hat{j}_{d - 1} \check{j}_{d - 1} ) ( \hat{j}_{d - 1}^{\prime} \check{j}_{d - 1}^{\prime} ) j_d j_d^{\prime}}.
  \end{equation}
  We can obtain $U_{( \hat{j}_1 \check{j}_1 ), ( \hat{k}_1 \check{k}_1 )}^{( 1 )}$ and $U_{( \hat{j}_1^{\prime} \check{j}_1^{\prime} ), ( \hat{k}_1^{\prime} \check{k}_1^{\prime} )}^{\prime ( 1 )}$ by the following singular value decomposition
  \begin{gather}
      \tilde{M}^{( 1 )} \left( \tilde{M}^{( 1 )} \right) ^{\dagger} = U^{( 1 )} \Lambda^{( 1 )} \left( U^{( 1 )} \right) ^{\dagger},  \\
      \tilde{M}^{\prime ( 1 )} \left( \tilde{M}^{\prime ( 1 )} \right) ^{\dagger} = U^{\prime ( 1 )} \Lambda^{\prime ( 1 )} \left( U^{\prime ( 1 )} \right) ^{\dagger}.
  \end{gather}
  Representations of matrices
  $\left( \tilde{M}^{( 1 )} \left( \tilde{M}^{( 1 )} \right) ^{\dagger} \right) _{( \hat{j}_1^{( \textrm{row} )} \check{j}_1^{( \textrm{row} )} ), ( \hat{j}_1^{( \textrm{column} )} \check{j}_1^{( \textrm{column} )} )}$
  and
  
\begin{figure}[H]
       \begin{center}
           \includegraphics[width=1.0\textwidth]{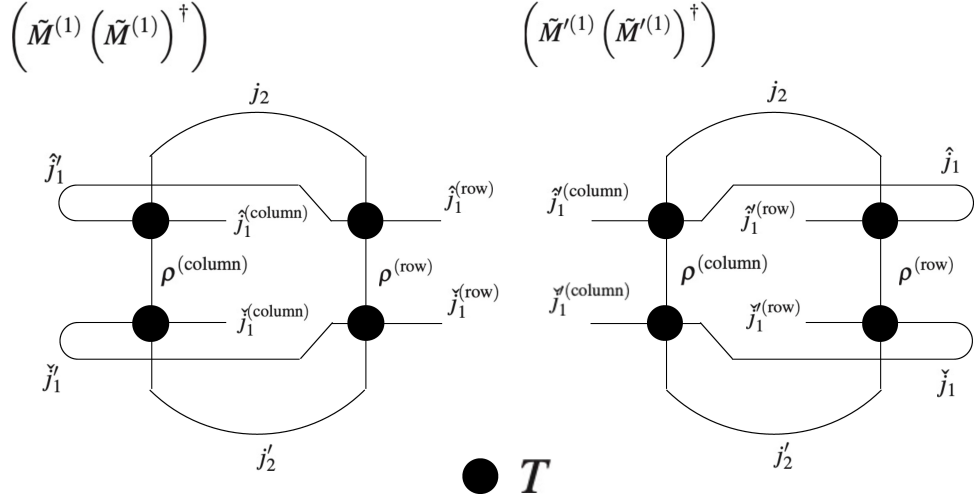}  \\
           \caption{Description of The matrices $\tilde{M}^{( 1 )} \left( \tilde{M}^{( 1 )} \right) ^{\dagger} $ and $\tilde{M}^{\prime ( 1 )} \left( \tilde{M}^{\prime ( 1 )} \right) ^{\dagger}$ to which SVD is applied as tensor network}
            \label{FIg_Matrix_SVD}
        \end{center}
    \end{figure}

  \noindent
  $\left( \tilde{M}^{\prime ( 1 )} \left( \tilde{M}^{\prime ( 1 )} \right) ^{\dagger} \right) _{( \hat{j}_1^{\prime ( \textrm{row} )} \check{j}_1^{\prime ( \textrm{row} )} ), ( \hat{j}_1^{\prime ( \textrm{column} )} \check{j}_1^{\prime ( \textrm{column} )} )}$ as tensor network
  are shown in Fig. \ref{FIg_Matrix_SVD}.
  \noindent
  Let us denote singular values of the matrices $\Lambda^{( 1 )}$ and $\Lambda^{^{\prime} ( 1 )}$ by $\sigma_{l}^{( 1 )}$ and $\sigma_{l}^{\prime ( 1 )}$ $( l = 1, ..., \chi_1^2 )$, respectively.
  These singular values are ordered in descending order, namely, $\sigma_1^{( 1 )} \geq \cdots \geq \sigma_{\chi_1^2}^{( 1 )}$ and $\sigma_1^{\prime ( 1 )} \geq \cdots \geq \sigma_{\chi_1^2}^{\prime ( 1 )}$.
  A way to choose one of the unitary matrices $U_{( \hat{j}_1 \check{j}_1 ), ( \hat{k}_1 \check{k}_1 )}^{( 1 )}$ and $U_{( \hat{j}_1^{\prime} \check{j}_1^{\prime} ), ( \hat{k}_1^{\prime} \check{k}_1^{\prime} )}^{\prime ( 1 )}$ is not unique.
  In \cite{XCQZYX12}, the following way is presented.
  Let us introduce the following quantities
  \begin{gather}
      \varepsilon^{( 1 )} = \sum_{l = \chi + 1}^{\chi_1^2} \sigma_{l}^{( 1 )},  \\
      \varepsilon^{\prime ( 1 )} = \sum_{l = \chi + 1}^{\chi_1^2} \sigma_{l}^{\prime ( 1 )}.
  \end{gather}
  If $\varepsilon^{( 1 )} < \varepsilon^{\prime ( 1 )}$, the unitary matrix $U_{( \hat{j}_1 \check{j}_1 ), ( \hat{k}_1 \check{k}_1 )}^{( 1 )}$ is adopted.
  If $\varepsilon^{( 1 )} \geq \varepsilon^{\prime ( 1 )}$, the unitary matrix $U_{( \hat{j}_1^{\prime} \check{j}_1^{\prime} ), ( \hat{k}_1^{\prime} \check{k}_1^{\prime} )}^{\prime ( 1 )}$ is adopted.

  Above mentioned coarse-graining procedure is applicable to the other directions in a similar way.

\begin{figure}[H]
       \begin{center}
           \includegraphics[width=0.3\textwidth]{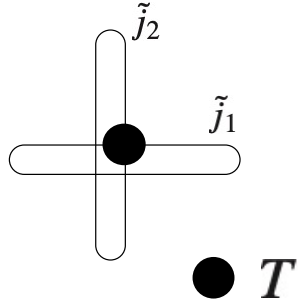}  \\
           \caption{Description of (\ref{Z_tTr}) as tensor network}
            \label{FIg_tTr_Z}
        \end{center}
    \end{figure}

  A lattice consists of $2^p$ sites is coarse-grained as one local tensor after $p$ times coarse-graining procedure.
  Assume that $p$ is sufficiently large and all the indices of the coarse-grained runs from $0, 1, ..., \chi - 1$ for specified bond dimension $\chi$.
  Then, we obtain approximated partition function $Z_{\text{approx}}$ of a lattice consists of $2^p$ sites with periodic boundary condition through a specified bond dimension $\chi$ as
  \begin{equation}  \label{Z_tTr}
      Z_{\text{approx}} = \textrm{tTr} T_{j_1 j_1^{\prime} ... j_d j_d^{\prime}} = \sum_{\tilde{j}_1 = 0}^{\chi - 1} \cdots \sum_{\tilde{j}_d = 0}^{\chi - 1} T_{\tilde{j}_1 \tilde{j}_1 ... \tilde{j}_d \tilde{j}_d}.
  \end{equation}
  Representation of (\ref{Z_tTr}) as tensor network is shown in Fig. \ref{FIg_tTr_Z}.

 \section{Key ideas of presented parallel computing method for the HOTRG}  \label{ParallelMethod}
 
  In this section, we describe key ideas of a parallel computing method for the HOTRG in a $d ( \geq 2 )$-dimensional simple lattice.
  In Section \ref{PreconditionImplement}, general principles in our method are given.
  Terminologies for our method and key ideas are explained.
  In Section \ref{Key_ideas}, we present key ideas of our method and explain how we conceive these ideas.

  \subsection{General principles in the presented method}  \label{PreconditionImplement}

  In this section, we give general principles in the presented method.
  They are explained in each subsections.
  
 \subsubsection{Usage of the word coarse-graining}
 
 Hereafter, the word coarse-graining means procedure of coarse-graining to one direction in this paper.

  \subsubsection{Identification of directions of a tensor network}  \label{Directions_TN}

\begin{figure}[H]
      \begin{center}
         \includegraphics[width=0.9\textwidth]{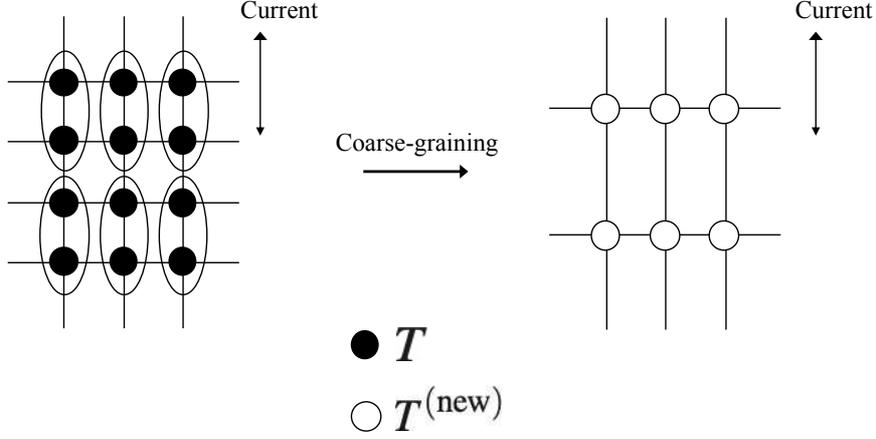}
         \caption{The Current direction}
         \label{Fig_Current}
      \end{center}
  \end{figure}

  Coarse-graining is applied to each direction alternately.
  Let us call the direction which we apply this procedure Current.
  See Fig. \ref{Fig_Current}.
  From two local tensors $T$ lined in the Current direction, a new local tensor $T^{(\textrm{new})}$ is constructed.
  Let us call the direction which we apply this procedure in the next coarse-graining Next 1.
  The directions Next 2,  Next 3 and so on are named in a similar manner.
  Of course, these directions are renamed in the next coarse-graining procedure.

  \subsubsection{Local tensors}  \label{LocalTensors}

  A local tensor is denoted by $T_{n_{d - 1} \dots n_1 n_{d - 1}^{\prime} \dots n_1^{\prime} c c^{\prime}}$.
  The indices $c$ and $c^{\prime}$ are ones in the direction Current.
  The indices $n_{j}$ and $n_{j}^{\prime}$ $( j = 1, ..., d - 1)$ are ones in the direction Next $j$.
  A new local tensor obtained through once coarse-graining procedure is denoted by $T_{n_{d-1} \dots n_1 n_{d-1}^{\prime} \dots n_1^{\prime} c c^{\prime}}^{(\textrm{new})}$.
 
  Let us denote bond dimension which is a parameter as truncation of unitary matrices by $\chi$.
  For a local tensor $T_{n_{d - 1} \dots n_1 n_{d - 1}^{\prime} \dots n_1^{\prime} c c^{\prime}}$,
  assume that the indices $c$ and $c^{\prime}$ take values $0$, $1$, ..., $\chi_c - 1$ $( \chi_c \leq \chi )$ and those $n_j$ and $n_j^{\prime}$ $( j = 1, ..., d - 1)$ take values $0$, $1$, ..., $\chi_{n_j} - 1$ $( \chi_{n_j} \leq \chi )$.
  In our method, local tensor elements are normalized by some factor $f^{( p )}$ before the $( p + 1 )$-th $( p = 0, 1, 2, ... )$ coarse-graining procedure to delay overflow.
  A way to determine the value of $f^{( p )}$ is not unique and $f^{( p )}$ may be different in each $p$.
  For example, one can use the inverse of the trace of a local tensor
  \begin{equation}  \label{Trace_LocalTensor}
      \textrm{Tr} T = \sum_{\tilde{n}_{d - 1} = 0}^{\chi_{d - 1} - 1} \cdots \sum_{\tilde{n}_1= 0}^{\chi_1 - 1}  \sum_{\tilde{c} = 0}^{\chi_c - 1} T_{\tilde{n}_{d-1} \dots \tilde{n}_1 \tilde{n}_{d-1} \dots \tilde{n}_1 \tilde{c} \tilde{c}}
  \end{equation}
  as a normalization factor under periodic boundary condition, that is, $f^{( p )} = ( \textrm{Tr} T )^{- 1}$.
  Before the first coarse-graining procedure, a local tensor $T_{n_{d - 1} \dots n_1 n_{d - 1}^{\prime} \dots n_1^{\prime} c c^{\prime}}^{( \textrm{init} )}$ constructed from a considering model is normalized as
  \begin{equation}
      T_{n_{d - 1} \dots n_1 n_{d - 1}^{\prime} \dots n_1^{\prime} c c^{\prime}} = f^{( 0 )} T_{n_{d - 1} \dots n_1 n_{d - 1}^{\prime} \dots n_1^{\prime} c c^{\prime}}^{( \textrm{init} )}
  \end{equation}
  and the first coarse-graining procedure is applied to this normalized tensor.
  After the $q$-th $( q = 1, 2, ... )$ coarse-graining procedure, obtained new local tensor is normalized and we have to consider renaming of indices of the new local tensor as mentioned in Section \ref{Directions_TN}.
  Then, in the $( q + 1 )$-th coarse-graining procedure, we apply coarse-graining procedure to the following local tensor
  \begin{equation}  \label{rename_indices}
      T_{\bar{n}_{d - 1} \dots \bar{n}_1 \bar{n}_{d - 1}^{\prime} \dots \bar{n}_1^{\prime} \bar{c} \bar{c}^{\prime}}^{(\textrm{next})} = f^{( q )} T_{n_{d-1} \dots n_1 n_{d-1}^{\prime} \dots n_1^{\prime} c c^{\prime}}^{(\textrm{new})},
  \end{equation}
  where the indices of the tensor $T_{\bar{n}_{d - 1} \dots \bar{n}_1 \bar{n}_{d - 1}^{\prime} \dots \bar{n}_1^{\prime} \bar{c} \bar{c}^{\prime}}^{(\textrm{next})}$ are
  \begin{gather}
      \bar{c} = n_1,   \\
      \bar{c}^{\prime} = n_1^{\prime},  \\
      \bar{n}_j = n_{j + 1},  \qquad  (j = 1, ..., d - 2),   \\
      \bar{n}_j^{\prime} = n_{j + 1}^{\prime},  \qquad  (j = 1, ..., d - 2),   \\
      \bar{n}_{d - 1} = c,   \\
      \bar{n}_{d - 1}^{\prime} = c^{\prime}.
  \end{gather}
  These indices take the following values
  \begin{gather*}
      \bar{c}, ~ \bar{c}^{\prime} : ~ 0, 1, ..., \chi_{n_1}^{(\textrm{new})} - 1,  \\
      \bar{n}_j, ~ \bar{n}_j^{\prime} : ~ 0, 1, ..., \chi_{n_{j + 1}}^{(\textrm{new})} - 1,  \qquad  ( j = 1 , ..., d - 2 ),  \\
      \bar{n}_{d - 1}, ~ \bar{n}_{d - 1}^{\prime} : ~ 0, 1, ..., \chi_c- 1,
  \end{gather*}
  where $\chi_{n_j}^{(\textrm{new})}$ $( j = 1 , ..., d - 1 )$ are $\chi_{n_j}^{(\textrm{new})} = \min \{ \chi_{n_j}^2, \chi \}$.

  \subsubsection{Two sides in each direction}  \label{Sides_in_Direction}

\begin{figure}[H]
      \begin{center}
         \includegraphics[width=0.7\textwidth]{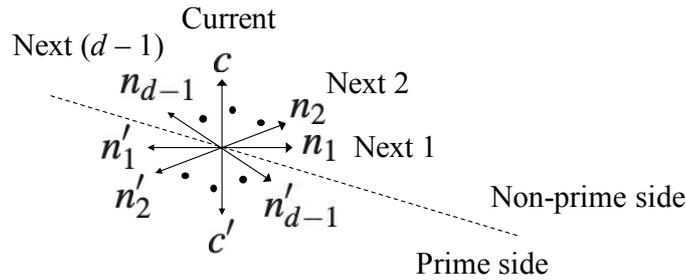}
         \caption{Two sides in each direction}
         \label{Fig_Sides}
      \end{center}
  \end{figure}

  As shown in Section \ref{LocalTensors}, a local tensor has two indices in each direction.
  It means that each direction has two sides.
  Let us call a side which is concerned with the index without prime Non-prime side.
  Similarly, let us call a side which is concerned with the index with prime Prime side.
  See Fig. \ref{Fig_Sides}.

  \subsubsection{Expression of processes in parallel computing}

  For simplicity, the process which has process number $P$ is expressed as the Process $P$.

  \subsection{Key ideas of the presented method}  \label{Key_ideas}

  In this section, we explain how to distribute elements of a tensor to each process.
  This way of distribution is the key ideas of our method.
  In parallel computing, the simplest way of distribution of local tensor elements to each process,
  say, one tensor element is placed to one process,
  causes a problem of cost for communication between processes in contraction procedure.
  This problem is caused by placement of necessary local tensor elements to more than one process.
  Then, to reduce communication between processes, it is natural to adopt the following principle:
  \begin{itemize}
    \item We accept placement of a local tensor element to more than one process.
  \end{itemize}
  The next question is how to distribute elements of a tensor to each process.
  Let us analyze equations in coarse-graining procedure.
  To obtain a new tensor, the following contraction is executed.
  \begin{align}
      T_{n_{d - 1} \dots n_1 n_{d - 1}^{\prime} \dots n_1^{\prime} c c^{\prime}}^{(\textrm{new})}
      =\sum &\left( \tilde{U}_{( \hat{n}_{d - 1}^{\prime} \check{n}_{d - 1}^{\prime} ), n_{d - 1}^{\prime}}^{( d - 1 )} \cdots \tilde{U}_{( \hat{n}_1^{\prime} \check{n}_1^{\prime} ), n_1^{\prime}}^{( 1 )} 
                            T_{\check{n}_{d - 1} \dots \check{n}_1 \check{n}_{d - 1}^{\prime} \dots \check{n}_1^{\prime} \rho c^{\prime}} \right.  \notag \\
                 &~~~ \times \left. T_{\hat{n}_{d - 1} \dots \hat{n}_1 \hat{n}_{d - 1}^{\prime} \dots \hat{n}_1^{\prime} c \rho}
                                      \tilde{U}_{( \hat{n}_1 \check{n}_1 ), n_1}^{( 1 )} \cdots \tilde{U}_{( \hat{n}_{d - 1} \check{n}_{d - 1} ), n_{d - 1}}^{( d - 1 )} \right) ,  \label{Contraction_NewTensor}
  \end{align}
  where $\tilde{U}_{( \hat{n}_j^{\prime} \check{n}_j^{\prime} ), n_j^{\prime}}^{( j )}$ and $\tilde{U}_{( \hat{n}_j \check{n}_j ), n_j}^{( j )}$ $( j = 1, ..., d - 1 )$ are unitary matrices.
  Subscripts $( \hat{n}_j^{\prime} \check{n}_j^{\prime} )$ and $( \hat{n}_j \check{n}_j )$ $( j = 1, ..., d - 1 )$ of the unitary matrices
  mean $\hat{n}_j^{\prime} + \check{n}_j^{\prime} \chi_{n_j}$ and $\hat{n}_j + \check{n}_j \chi_{n_j}$, respectively.
  Summation $\sum$ means that
  \begin{equation}
     \sum_{\check{n}_{d - 1}^{\prime} = 0}^{\chi_{n_{d - 1}} - 1}
     \cdots
     \sum_{\check{n}_1^{\prime} = 0}^{\chi_{n_1} - 1}
     \sum_{\hat{n}_{d - 1}^{\prime} = 0}^{\chi_{n_{d - 1}} - 1}
     \cdots
     \sum_{\hat{n}_1^{\prime} = 0}^{\chi_{n_1} - 1}
     \sum_{\check{n}_{d - 1} = 0}^{\chi_{n_{d - 1}} - 1}
     \cdots
     \sum_{\check{n}_1 = 0} ^{\chi_{n_1} - 1}
     \sum_{\hat{n}_{d - 1} = 0}^{\chi_{n_{d - 1}} - 1}
     \cdots
     \sum_{\hat{n}_1 = 0}^{\chi_{n_1} - 1}
     \sum_{\rho = 0}^{\chi_c - 1}.
  \end{equation}
  The unitary matrices $\tilde{U}^{( j )}$ $( j = 1, ..., d - 1 )$ are obtained through SVD.
  In the Non-prime sides, SVD is given as
  \begin{equation}
     A^{( j )} = U^{( j )} \Sigma^{( j )} ( V^{( j )} )^{\top},  \qquad  ( j = 1, ..., d - 1 ).
  \end{equation}
  In the Prime sides, SVD is given as
  \begin{equation}
     A^{\prime ( j )} = U^{\prime ( j )} \Sigma^{\prime ( j )} ( V^{\prime ( j )} )^{\top},  \qquad  ( j = 1, ..., d - 1 ).
  \end{equation}
  The matrices $A^{( j )}$ and $A^{\prime ( j )}$ to which SVD is applied are obtained through the following three steps.
  Firstly, we consider the Non-prime side.
  In the first step, a tensor $\hat{S}_{\hat{f} \hat{g} \rho \eta}^{( j )}$ is computed as
  \begin{equation}  \label{Aux_Tensor_hat_S}
      \hat{S}_{\hat{f} \hat{g} \rho \eta}^{( j )}
      = \sum T_{n_{d-1} \dots n_{j + 1} \hat{f}     n_{j - 1} \dots n_1 n_{d-1}^{\prime} \dots n_1^{\prime} c \rho}
                  T_{n_{d-1} \dots n_{j + 1} \hat{g} n_{j - 1} \dots n_1 n_{d-1}^{\prime} \dots n_1^{\prime} c \eta},
  \end{equation}
  where $\sum$ means that
  \begin{equation}  \label{Aux_Tensor_Sum}
      \sum = \sum_{n_{d - 1} = 0}^{\chi_{n_{d - 1}} - 1} \cdots \sum_{n_{j + 1} = 0}^{\chi_{n_{j + 1} } - 1}
                  \sum_{n_{j - 1} = 0}^{\chi_{n_{j - 1}} - 1} \cdots \sum_{n_1 = 0}^{\chi_{n_1} - 1}
                  \sum_{n_{d - 1}^{\prime} = 0}^{\chi_{n_{d - 1}} - 1} \cdots \sum_{n_1^{\prime} = 0}^{\chi_{n_1} - 1}
                  \sum_{c = 0}^{\chi_c - 1}.
  \end{equation}
  In the second step, a tensor $\check{S}_{\check{f} \check{g} \rho \eta}^{( j )}$ is computed as
  \begin{equation}  \label{Aux_Tensor_check_S}
      \check{S}_{\check{f} \check{g} \rho \eta}^{( j )}
      = \sum T_{n_{d-1} \dots n_{j + 1} \check{f}     n_{j - 1} \dots n_1 n_{d-1}^{\prime} \dots n_1^{\prime} \rho c^{\prime} }
                  T_{n_{d-1} \dots n_{j + 1} \check{g} n_{j - 1} \dots n_1 n_{d-1}^{\prime} \dots n_1^{\prime} \eta c^{\prime} },  \\
  \end{equation}
  where $\sum$ is 
  \begin{equation}
      \sum = \sum_{n_{d - 1} = 0}^{\chi_{n_{d - 1}} - 1} \cdots \sum_{n_{j + 1} = 0}^{\chi_{n_{j + 1} } - 1}
                  \sum_{n_{j - 1} = 0}^{\chi_{n_{j - 1}} - 1} \cdots \sum_{n_1 = 0}^{\chi_{n_1} - 1}
                  \sum_{n_{d - 1}^{\prime} = 0}^{\chi_{n_{d - 1}} - 1} \cdots \sum_{n_1^{\prime} = 0}^{\chi_{n_1} - 1}
                  \sum_{c^{\prime} = 0}^{\chi_c - 1}.
  \end{equation}
  In the last step, the matrix $A^{( j )}$ is computed as
  \begin{equation} 
      A_{( \hat{f} \check{f} ), ( \hat{g} \check{g} )}^{( j )}
      =\sum_{\rho = 0}^{\chi_c - 1} \sum_{\eta = 0}^{\chi_c - 1} \hat{S}_{\hat{f} \hat{g} \rho \eta}^{( j )} \check{S}_{\check{f} \check{g} \rho \eta}^{( j )},
  \end{equation}
  where $( \hat{f} \check{f} )$ is $( \hat{f} \check{f} ) = \hat{f}  + \check{f} \chi_{n_j}$ and $( \hat{g} \check{g} )$ is $( \hat{g} \check{g} ) = \hat{g} + \check{g} \chi_{n_j}$.
  In the Prime sides, tensors $\hat{S}_{\hat{f} \hat{g} \rho \eta}^{\prime ( j )}$ and $ \check{S}_{\check{f} \check{g} \rho \eta}^{\prime ( j )}$ $( j = 1, ..., d - 1 )$ are introduced and computed in a similar way.
  The matrices $A^{\prime ( j )}$ are computed as
  \begin{equation} 
      A_{( \hat{f} \check{f} ), ( \hat{g} \check{g} )}^{\prime ( j )}
      =\sum_{\rho = 0}^{\chi_c - 1} \sum_{\eta = 0}^{\chi_c - 1} \hat{S}_{\hat{f} \hat{g} \rho \eta}^{\prime ( j )} \check{S}_{\check{f} \check{g} \rho \eta}^{\prime ( j )}.
  \end{equation}
  From Eqs. (\ref{Contraction_NewTensor}), (\ref{Aux_Tensor_hat_S}) and $(\ref{Aux_Tensor_check_S})$,
  we notice that all the elements of the local tensor $T$ are not necessary
  to compute each element of $T_{n_{d - 1} \dots n_1 n_{d - 1}^{\prime} \dots n_1^{\prime} c c^{\prime}}^{(\textrm{new})}$,
  $\hat{S}_{\hat{f} \hat{g} \rho \eta}^{( j )}$ and $\check{S}_{\check{f} \check{g} \rho \eta}^{( j )}$ $( j = 1, ..., d - 1 )$.
  It is similar to $\hat{S}_{\hat{f} \hat{g} \rho \eta}^{\prime ( j )}$ and $ \check{S}_{\check{f} \check{g} \rho \eta}^{\prime ( j )}$ $( j = 1, ..., d - 1 )$.
  It indicates that when we compute the elements of these tensors in parallel computing, we have only to store not all but sufficient elements in each process.
  Under  such distribution of elements, communication between processes does not occur during considering computation.
  Let us introduce $c$-fixed and $c^{\prime}$-fixed tensors.
  For a fixed $c ( = c_0 )$, the $c$-fixed tensor is defined as
  \begin{equation}
      \hat{T}_{n_{d-1} \dots n_1 n_{d-1}^{\prime} \dots n_1^{\prime} c^{\prime}}^{c = c_0} = T_{n_{d-1} \dots n_1 n_{d-1}^{\prime} \dots n_1^{\prime} c_0 c^{\prime}}.
  \end{equation}
  For a fixed $c^{\prime} ( = c_0^{\prime} )$, the $c^{\prime}$-fixed tensor is defined as
  \begin{equation}
     \check{T}_{n_{d-1} \dots n_1 n_{d-1}^{\prime} \dots n_1^{\prime} c}^{c^{\prime} = c_0^{\prime}} = T_{n_{d-1} \dots n_1 n_{d-1}^{\prime} \dots n_1^{\prime} c c_0^{\prime}},
  \end{equation}
  For fixed $c = c_0$ and $c^{\prime} = c_0^{\prime}$, we can compute the elements $T_{n_{d - 1} \dots n_1 n_{d - 1}^{\prime} \dots n_1^{\prime} c c^{\prime}}^{(\textrm{new})}$ in (\ref{Contraction_NewTensor})
                                                                                       from a $c$-fixed tensor $\hat{T}_{n_{d-1} \dots n_1 n_{d-1}^{\prime} \dots n_1^{\prime} c^{\prime}}^{c = c_0}$
                                                                                               and a $c^{\prime}$-fixed tensor $\check{T}_{n_{d-1} \dots n_1 n_{d-1}^{\prime} \dots n_1^{\prime} c}^{c^{\prime} = c_0^{\prime}}$.
  For fixed $\rho= \rho_0$ and $\eta = \eta_0$, we can compute the elements $\hat{S}_{\hat{f} \hat{g} \rho_0 \eta_0}^{( j )}$ and $\hat{S}_{\hat{f} \hat{g} \rho_0 \eta_0}^{\prime ( j )}$ $( j = 1, ..., d - 1 )$
                                                                           from $c^{\prime}$-fixed tensors $\check{T}_{n_{d-1} \dots n_1 n_{d-1}^{\prime} \dots n_1^{\prime} c}^{c^{\prime} = \rho_0}$
                                                                                    and $\check{T}_{n_{d-1} \dots n_1 n_{d-1}^{\prime} \dots n_1^{\prime} c}^{c^{\prime} = \eta_0}$.
  Similarly, we can compute the elements $\check{S}_{\check{f} \check{g} \rho_0 \eta_0}^{( j )}$ and $\check{S}_{\check{f} \check{g} \rho_0 \eta_0}^{\prime ( j )}$ $( j = 1, ..., d - 1 )$
                 from $c$-fixed tensors  $\hat{T}_{n_{d-1} \dots n_1 n_{d-1}^{\prime} \dots n_1^{\prime} c^{\prime}}^{c = \rho_0}$ and $\hat{T}_{n_{d-1} \dots n_1 n_{d-1}^{\prime} \dots n_1^{\prime} c^{\prime}}^{c =\eta_0}$.
  Thus, we derive the following principal on distribution of local tensor elements in parallel computing.
  \begin{itemize}
    \item Elements of a local tensor are distributed to each process according to one of indices.
  \end{itemize}
  Since two newly introduced tensors are distributed to each process, we prepare $\chi^2$ processes for parallel computing.
  The fixed indices are identified through a process number of each process in parallel computing.
  For example, in computation of (\ref{Contraction_NewTensor}), the Process $( c_0 + c_0^{\prime} \chi )$ has elements of the $c$-fixed tensor $\hat{T}_{n_{d-1} \dots n_1 n_{d-1}^{\prime} \dots n_1^{\prime} c^{\prime}}^{c = c_0}$
                                                                                                       and the $c^{\prime}$-fixed tensor $\check{T}_{n_{d-1} \dots n_1 n_{d-1}^{\prime} \dots n_1^{\prime} c}^{c^{\prime} = c_0^{\prime}}$.
  Note that the indices identified through a process number are not contracted during considering computation.
  Consequently, the key ideas are summarized as follows.
  
  At the beginning of some of steps  in the presented method, sufficient local tensor elements for computation are placed to each process to avoid communication between process during the step.
  We accept that an element of a local tensor is placed to more than one process.
  The index $c$ of a $c$-fixed tensor $\hat{T}_{n_{d-1} \dots n_1 n_{d-1}^{\prime} \dots n_1^{\prime} c^{\prime}}^{c = c_0}$
  and that $c^{\prime}$ of a $c^{\prime}$-fixed tensor $\check{T}_{n_{d-1} \dots n_1 n_{d-1}^{\prime} \dots n_1^{\prime} c}^{c^{\prime} = c_0^{\prime}}$
  which are not contracted during considering step
  are identified through process number of a process in parallel computing and distribution of local tensor elements is done according to these indices.

  If a method different from the HOTRG has a suitable mathematical structure, these ideas can be applicable to it.

 \section{Implementation of the presented method}  \label{Implementation}

  In this section, we give a way to implement our method in detail.
  In this section, we assume periodic boundary condition to a lattice.
  For dimensionality $d$ of a simple lattice and bond dimension $\chi$, when $d$ is $d \geq 3$,
  we can verify that computational cost in each process and memory space requirement in each process are $O ( \chi^{4d - 3} )$ and $O ( \chi^{2d - 1} )$, respectively, from this implementation.
  In the case of $d = 2$, a step for SVD is dominant in computational cost and memory space requirement.
  Computational cost in each process is $O ( \chi^6 )$ and memory space requirement in each process is $O ( \chi^4 )$.

  In Section \ref{PhysicalQuantities}, physical quantities computed in our method are explained before we describe the implementation.
  In Section \ref{PresentMethod}, we present an implementation of our method.

  \subsection{Physical quantities in the presented method}  \label{PhysicalQuantities}

  In this section, we mention physical quantities which are computed in the presented method.
  They are the partition function and a ratio \cite{GW09} which can be used to identify a phase of a model.

  In Section \ref{PartitionFunction}, we mention the partition function $Z$.
  For physics, quantity $\ln Z / V$, where $V$ is the volume of a lattice, is important.
  Since normalization of a local tensor is done in our method, the details of approximation of the quantity $\ln Z / V$ is explained in this section.
  In Section \ref{RatioForTc}, the ratio is mentioned.

 \subsubsection{The partition function}  \label{PartitionFunction}
 
\begin{figure}[H]
      \begin{center}
         \includegraphics[width=0.4\textwidth]{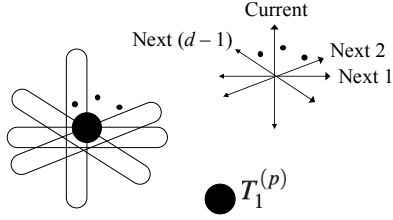}
         \caption{Representation of the trace of the local tensor $T_1^{(p)}$ as a tensor network}
         \label{Fig_TensorTrace}
      \end{center}
  \end{figure}

  In this section, we explain approximation of the quantity $\ln Z / V$, where $Z$ and $V$ are the partition function and the volume of a lattice, respectively.
  Let us consider a lattice which has $V = V^{( p )} = 2^p$ lattice points with periodic boundary condition, where $p$ is a positive integer.
  The partition function is given as
  \begin{equation}
      Z = \sum \left( \prod_{i = 1}^{V^{( p )}} T_i^{(0)} \right) ,
  \end{equation}
  where $T_i^{(0)}$ is the local tensor in lattice point $i$ and summation is taken for all the cases of configuration of indices of local tensors.
  As mentioned in Section \ref{LocalTensors}, local tensor elements are normalized by some factor before procedure of each coarse-graining 
  and effect of this normalization should be considered in computation of the partition function.
  Assume that the local tensors $T_i^{(0)}$ are normalized as
  \begin{equation}  \label{Normalization_Initial}
      \breve{T}_i^{(0)} = f^{( 0 )} T_i^{(0)}.
  \end{equation}
  Then, partition function is expressed as
  \begin{equation}
      Z  = \sum \left( \prod_{i = 1}^{V^{( p )}} ( f^{( 0 )} )^{- 1} \breve{T}_i^{(0)} \right)  = ( f^{( 0 )} )^{- V^{( p )}} \sum \left( \prod_{i = 1}^{V^{( p )}}  \breve{T}_i^{(0)} \right) .
  \end{equation}
  After once coarse-graining, we have new local tensors.
  Let us denote these new local tensors by $T_i^{(1)}$.
  Partition function is approximately given as
  \begin{equation}
      \tilde{Z}^{( p, 1 )}  = ( f^{( 0 )} )^{- V^{( p )}} \sum \left( \prod_{i = 1}^{V^{( p )} / 2}  T_i^{(1)} \right) 
  \end{equation}
  because of truncation of unitary matrices.
  Assume that local tensors $T_i^{(1)}$ are normalized as $\breve{T}_i^{(1)} = f^{( 1 )} T_i^{(1)}$.
  Then, the partition function is approximately
  \begin{align}
      \tilde{Z}^{( p, 1 )} &= ( f^{( 0 )} )^{- V^{( p )}} \sum \left( \prod_{i = 1}^{V^{( p )} / 2} ( f^{( 1 )} )^{- 1} \breve{T}_i^{(1)} \right)  \notag \\
                                  &= ( f^{( 0 )} )^{- V^{( p )}} ( f^{( 1 )} )^{- V^{( p )} / 2} \sum \left( \prod_{i = 1}^{V^{( p )} / 2} \breve{T}_i^{(1)} \right) .
  \end{align}
  Repeating this procedure, we have approximated partition function as
  \begin{equation}  \label{Z_OneTensor} 
      Z^{( p )} = \tilde{Z}^{( p, p )}  = \left( \prod_{l= 0}^{p - 1} ( f^{( l )} )^{- V^{( p )} / 2^l} \right) \cdot \sum T_1^{(p)}.
  \end{equation}
  Because of periodic boundary condition, it holds
  \begin{equation}
      \sum T_1^{(p)} = \textrm{Tr} T_1^{(p)}
                               = \sum_{\tilde{n}_{d - 1} = 0}^{\chi_{d - 1} - 1} \cdots \sum_{\tilde{n}_1= 0}^{\chi_1 - 1}  \sum_{\tilde{c} = 0}^{\chi_c - 1} \left( T_1^{(p)} \right) _{\tilde{n}_{d - 1} \dots \tilde{n}_1 \tilde{n}_{d - 1} \dots \tilde{n}_1 \tilde{c} \tilde{c}}.
  \end{equation}
  Namely, the term $\sum T_1^{(p)}$ is equal to the trace of the local tensor $T_1^{(p)}$.
  Representation of this trace as a tensor network is shown in Fig. \ref{Fig_TensorTrace}.
  \noindent
  Thus, the partition function is approximately given as
  \begin{equation}
      Z^{(p)} = \left( \prod_{l= 0}^{p - 1} ( f^{( l )} )^{- 2^{p - l}} \right) \cdot \textrm{Tr} T_1^{(p)}.
  \end{equation}
  Taking the logarithm of the both sides, we have
  \begin{equation}  \label{eq_lnZ}
      \ln Z^{(p)} =  K^{( p )}  + \ln \left( \textrm{Tr} T_1^{(p)} \right) ,  \qquad  \left( K^{( p )}  = \sum_{l= 0}^{p - 1} 2^{p - l} \ln \left( f^{( l )} \right) ^{- 1} \right) .
  \end{equation}
  The term $K^{( p )} $ in (\ref{eq_lnZ}) can be computed through a recurrence relation.
  It holds
  \begin{equation}
      \left \{ \begin{array}{l}
                   K^{( 0 )} = 0,    \\
                   K^{( q )} = 2 \left( K^{( q - 1 )} + \ln \left( f^{( q - 1 )} \right) ^{- 1}  \right) , \qquad  ( q = 1, 2, ... ) . 
                \end{array} \right.   \label{K_recurrence}
  \end{equation}
  Thus, we can obtain approximation of the quantity $\ln Z / V$ as $\ln Z^{( p )} / V^{( p )}$, where $V = V^{( p )} = 2^p$.

 \subsubsection{A ratio to identify a phase of a model}  \label{RatioForTc}

\begin{figure}[H]
      \begin{center}
         \includegraphics[width=0.4\textwidth]{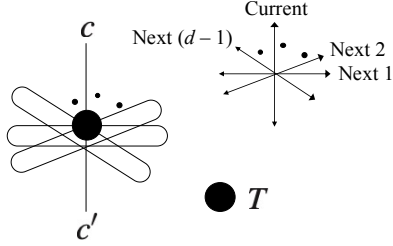}
         \caption{Tensor network representation of matrix $Y$ in (\ref{Matrix_Y})}
         \label{Fig_MatForTr}
      \end{center}
  \end{figure}

  Let us consider phase transition between a disordered phase and a symmetry breaking phase with $m$ degenerate states.
  To know which phase appears under a given condition, we can use a ratio introduced by Gu and Wen \cite{GW09}.
  This ratio is given as
  \begin{equation}  \label{ratio_X}
      X = \frac{( \textrm{trace} Y )^2}{\textrm{trace} ( Y^2 )},
  \end{equation}
  where $Y = ( Y_{c, c^{\prime}} )$ $( Y \in \mathbb{R}^{\chi_c \times \chi_c} )$ is a matrix defined as
  \begin{equation}  \label{Matrix_Y}
      Y_{c, c^{\prime}} =  \sum_{\tilde{n}_{d - 1} = 0}^{\chi_{d - 1} - 1}  \cdots \sum_{\tilde{n}_1 = 0}^{\chi_1 - 1} T_{\tilde{n}_{d-1} \dots \tilde{n}_1 \tilde{n}_{d-1} \dots \tilde{n}_1 c c^{\prime}}.
  \end{equation}
  Tensor network representation of the matrix $Y = ( Y_{c, c^{\prime}} )$ is shown in Fig. \ref{Fig_MatForTr}.
  For a disordered phase (a symmetry breaking phase with $m$ degenerate states), this ratio $X$ theoretically converges to $1$ ($m$) as coarse-graining procedure is iterated.
  From this ratio, we can estimate a range in which the critical point exists.
  For details, see \cite{GW09}.

  \subsection{Presentation of the method}  \label{PresentMethod}

  In this section, we present our method.
  The procedures described in this section are executed in all the processes in parallel computing unless otherwise noted.
  Section \ref{Prep_CG} is the procedure before the first coarse-graining.
  Sections from \ref{StateBeginning} to \ref{Normalization_NewTensors} are descriptions of once coarse-graining procedure.

  \subsubsection{Preparation for coarse-graining}  \label{Prep_CG}

  In this section, procedures before the first coarse-graining is given.

  Value of $K^{( 0 )}$ given in (\ref{K_recurrence}) is set to zero.
  
  The bond dimensions $\chi_c$ and $\chi_{n_j}$ $( j = 1, ..., d - 1)$ are set according to considering model.
  New bond dimensions $\chi_{n_j}^{(\textrm{new})}$ $( j = 1, ..., d - 1)$ of the new local tensor obtained after the first coarse-graining are computed by $\chi_{n_j}^{(\textrm{new})} = \min \{ \chi_{n_j}^2, \chi \}$.
  
  On the initial local tensor, the $c$-fixed tensor $\hat{T}_{n_{d-1} \dots n_1 n_{d-1}^{\prime} \dots n_1^{\prime} c^{\prime}}^{c = c_0}$ is set in the Process $c_0 + c_0 \chi$
                                           and the $c^{\prime}$-fixed tensor $\check{T}_{n_{d-1} \dots n_1 n_{d-1}^{\prime} \dots n_1^{\prime} c}^{c^{\prime} = c_0^{\prime}} $ is set in the Process $c_0 ^{\prime}+ c_0^{\prime} \chi$.

  Since we observe transition of quantities $\ln Z^{( p )} / V^{( p )}$ $( p = 0, 1, 2, ... )$ and the quantity $\ln Z^{( 0 )} / V^{( 0 )}$ is the trace of the initial local tensor from (\ref{eq_lnZ}) and (\ref{K_recurrence}),
  we compute the trace of the initial local by the following procedure.
  A variable $v_{c_0, c_0^{\prime}}$ is set to zero.
  In the Processes $s + s \chi$ $( s = 0, 1, ..., \chi_c - 1 )$, namely, processes in which the $c$-fixed tensor is computed, we execute the following computation
  \begin{equation}
      v_{s, s} = \sum_{\tilde{n}_{d - 1} = 0}^{\chi_{d - 1} - 1} \cdots \sum_{\tilde{n}_1 = 0}^{\chi_1 - 1} \hat{T}_{\tilde{n}_{d - 1} \dots \tilde{n}_1 \tilde{n}_{d - 1} \dots \tilde{n}_1 s}^{c = s}.
  \end{equation}
  Summation of $v_{c_0, c_0^{\prime}}$ is taken over all the processes since the trace is given as
  \begin{equation}
      \textrm{Tr} T_1^{( 0 )} = \sum_{c_0 = 0}^{\chi - 1} \sum_{c_0^{\prime} = 0}^{\chi - 1} v_{c_0, c_0^{\prime}}
                                         = \sum_{s= 0}^{\chi_c - 1} v_{s, s}
  \end{equation}
  and its result is shared among all the processes.
  In these procedures, summation and sharing, communication between processes occurs.

  We compute a normalization factor $f^{( 0 )}$ in some way.
  Substituting this factor into (\ref{K_recurrence}), we have the quantity $K^{( 1 )}$.
  The initial $c$-fixed and the $c^{\prime}$-fixed tensors are normalized as
  \begin{gather}
      \hat{T}_{n_{d-1} \dots n_1 n_{d-1}^{\prime} \dots n_1^{\prime} c^{\prime}}^{c = c_0, \textrm{Normalized}} = f^{( 0 )} \hat{T}_{n_{d-1} \dots n_1 n_{d-1}^{\prime} \dots n_1^{\prime} c^{\prime}}^{c = c_0},  \\
      \check{T}_{n_{d-1} \dots n_1 n_{d-1}^{\prime} \dots n_1^{\prime} c}^{c^{\prime} = c_0^{\prime}, \textrm{Normalized}} = f^{( 0 )} \check{T}_{n_{d-1} \dots n_1 n_{d-1}^{\prime} \dots n_1^{\prime} c}^{c^{\prime} = c_0^{\prime}},
  \end{gather}
  respectively.
  Then, the tensors $\hat{T}_{n_{d-1} \dots n_1 n_{d-1}^{\prime} \dots n_1^{\prime} c^{\prime}}^{c = c_0, \textrm{Normalized}}$ and $\check{T}_{n_{d-1} \dots n_1 n_{d-1}^{\prime} \dots n_1^{\prime} c}^{c^{\prime} = c_0^{\prime}, \textrm{Normalized}}$
  are newly regarded as the $c$-fixed and the $c^{\prime}$-fixed tensors $\hat{T}_{n_{d-1} \dots n_1 n_{d-1}^{\prime} \dots n_1^{\prime} c^{\prime}}^{c = c_0}$
                                      and $\check{T}_{n_{d-1} \dots n_1 n_{d-1}^{\prime} \dots n_1^{\prime} c}^{c^{\prime} = c_0^{\prime}}$, respectively,
  and coarse-graining procedure is applied to them.

  \subsubsection{State at the beginning of each coarse-graining procedure}  \label{StateBeginning}

  At the beginning of each coarse-graining, the $c$-fixed tensor $\hat{T}_{n_{d-1} \dots n_1 n_{d-1}^{\prime} \dots n_1^{\prime} c^{\prime}}^{c = c_0}$
                                                                    and the $c^{\prime}$-fixed tensor $\check{T}_{n_{d-1} \dots n_1 n_{d-1}^{\prime} \dots n_1^{\prime} c}^{c^{\prime} = c_0^{\prime}}$
  are stored in the Process $( c_0 + c_0 \chi )$ and  the Process $( c_0^{\prime} + c_0^{\prime} \chi )$, respectively.

  \subsubsection{Broadcasting of  the $c$-fixed and the $c^{\prime}$-fixed tensors to each process}  \label{BroadcastBeginning}

\begin{figure}[H]
      \begin{center}
         \includegraphics[width=0.7\textwidth]{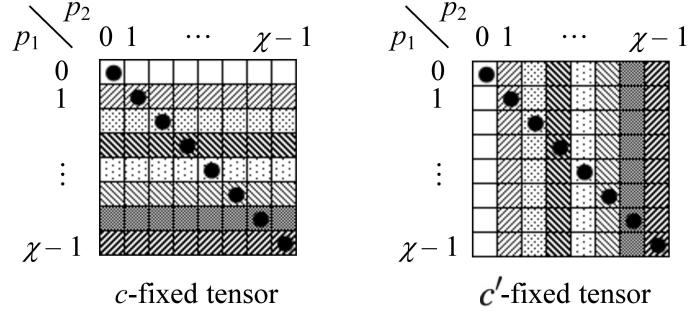}
         \caption{Image of broadcasting of the tensors}
         \label{Fig_BcastBegin}
      \end{center}
  \end{figure}

  The $c$-fixed tensor in the Process $( p_1 + p_1 \chi )$ is broadcasted to the Processes $( p_1 + \tilde{p} \chi )$ $( \tilde{p} = 0, 1, ..., \chi - 1 )$ except the Process $( p_1 + p_1 \chi )$ itself.
  Similarly, the $c^{\prime}$-fixed tensor in the Process $( p_2 + p_2 \chi )$ is broadcasted to the Processes $( \tilde{p} + p_2 \chi )$ $( \tilde{p} = 0, 1, ..., \chi - 1 )$ except the Process $( p_2 + p_2 \chi )$ itself.
  After broadcasting, the Process $( p_1 + p_2 \chi )$ has the $c$-fixed tensor $\hat{T}_{n_{d-1} \dots n_1 n_{d-1}^{\prime} \dots n_1^{\prime} c^{\prime}}^{c = p_1}$
  and the $c^{\prime}$-fixed tensor $\check{T}_{n_{d-1} \dots n_1 n_{d-1}^{\prime} \dots n_1^{\prime} c}^{c^{\prime} = p_2}$.
  See Fig. \ref{Fig_BcastBegin} for help of understanding.
  In this figure, each square represents each process.
  Each row and column represent values $p_1$ and $p_2$ $( = 0, 1, ..., \chi - 1 )$, respectively.
  The box in the intersection of row $p_1$ and column $p_2$ represents the Process $( p_1 + p_2 \chi )$.
  The $c$-fixed and the $c^{\prime}$-fixed tensors are broadcasted to horizontal and vertical directions, respectively.

  \subsubsection{Construction of matrices to which SVD is applied}  \label{MatricesForSVD}

  For dimensionality $d$ of a lattice, $2 ( d - 1 )$ unitary matrices are necessary to compute a new tensor through coarse-graining.
  These unitary matrices are obtained through singular value decomposition (SVD) of a matrix.
  In this section, we describe a method to compute the matrices to which SVD is applied.
  These matrices are computed through the three steps shown in Section \ref{Key_ideas}.

\begin{figure}[H]
      \begin{center}
         \includegraphics[width=0.9\textwidth]{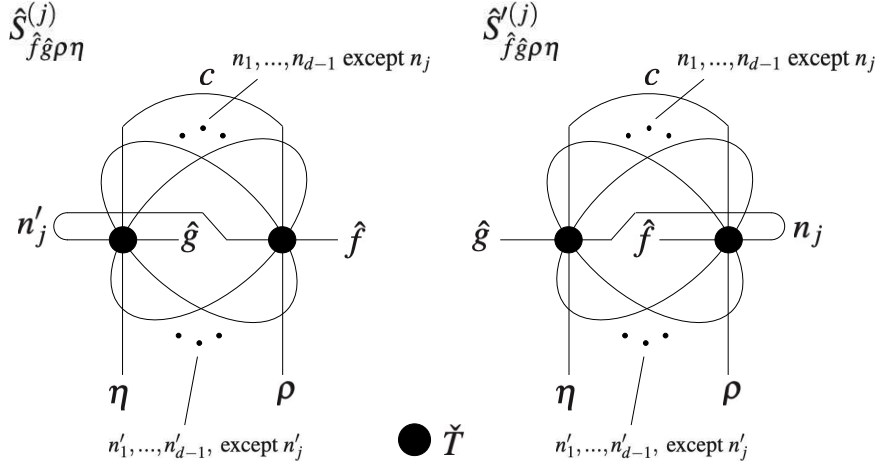}
         \caption{Representation of tensors $\hat{S}_{\hat{f} \hat{g} \rho \eta}^{( j )}$ and $\hat{S}_{\hat{f} \hat{g} \rho \eta}^{\prime ( j )}$ by tensor network}
         \label{Fig_TensorsS_hat}
      \end{center}
  \end{figure}

  In the first step, for $j = 1, ..., d - 1$, the following tensors are computed from two $c^{\prime}$-fixed tensors.
  \begin{gather}
      \hat{S}_{\hat{f} \hat{g} \rho \eta}^{( j )}
      = \sum \check{T}_{n_{d-1} \dots n_{j + 1} \hat{f}     n_{j - 1} \dots n_1 n_{d-1}^{\prime} \dots n_1^{\prime} c}^{c^{\prime} = \rho}
                  \check{T}_{n_{d-1} \dots n_{j + 1} \hat{g} n_{j - 1} \dots n_1 n_{d-1}^{\prime} \dots n_1^{\prime} c}^{c^{\prime} = \eta},   \label{def_hat_S_N} \\
      \hat{S}_{\hat{f} \hat{g} \rho \eta}^{\prime ( j )}
      = \sum \nolimits' \check{T}_{n_{d-1} \dots n_1 n_{d-1}^{\prime} \dots n_{j + 1}^{\prime} \hat{f}     n_{j - 1}^{\prime} \dots n_1^{\prime} c}^{c^{\prime} = \rho}
                                 \check{T}_{n_{d-1} \dots n_1 n_{d-1}^{\prime} \dots n_{j + 1}^{\prime} \hat{g} n_{j - 1}^{\prime} \dots n_1^{\prime} c}^{c^{\prime} = \eta},  \label{def_hat_S_P}
  \end{gather}
  where $\sum$ and $\sum^{\prime}$ are
  \begin{equation}  \label{contraction_hat_S_N}
      \sum = \sum_{n_{d - 1} = 0}^{\chi_{n_{d - 1}} - 1} \cdots \sum_{n_{j + 1} = 0}^{\chi_{n_{j + 1} } - 1}
                  \sum_{n_{j - 1} = 0}^{\chi_{n_{j - 1}} - 1} \cdots \sum_{n_1 = 0}^{\chi_{n_1} - 1}
                  \sum_{n_{d - 1}^{\prime} = 0}^{\chi_{n_{d - 1}} - 1} \cdots \sum_{n_1^{\prime} = 0}^{\chi_{n_1} - 1}
                  \sum_{c = 0}^{\chi_c - 1}
  \end{equation}
  and
  \begin{equation} \label{contraction_hat_S_P}
      \sum \nolimits' = \sum_{n_{d - 1} = 0}^{\chi_{n_{d - 1}} - 1} \cdots \sum_{n_1 = 0}^{\chi_{n_1} - 1}
                                \sum_{n_{d - 1}^{\prime} = 0}^{\chi_{n_{d - 1}} - 1} \cdots \sum_{n_{j + 1}^{\prime} = 0}^{\chi_{n_{j + 1} } - 1}
                                \sum_{n_{j - 1}^{\prime} = 0}^{\chi_{n_{j - 1}} - 1} \cdots \sum_{n_1^{\prime} = 0}^{\chi_{n_1} - 1}
                                \sum_{c = 0}^{\chi_c - 1},
  \end{equation}
  respectively.
  Their representations by tensor network is shown in Fig. \ref{Fig_TensorsS_hat}.
  These representations correspond to upper half of tensor network shown in Fig. \ref{FIg_Matrix_SVD}.
  In our method, The elements of $\hat{S}_{\hat{f} \hat{g} \rho \eta}^{( j )}$ and $\hat{S}_{\hat{f} \hat{g} \rho \eta}^{\prime ( j )}$ are computed in the Process $( \rho + \eta \chi )$.
  The $c^{\prime}$-fixed tensor $\check{T}_{n_{d-1} \dots n_1 n_{d-1}^{\prime} \dots n_1^{\prime} c}^{c^{\prime} = \eta}$ have already been stored in each process by broadcasting explained in Section \ref{BroadcastBeginning}.
  The $c^{\prime}$-fixed tensor in the Process $( p_1 + p_1 \chi )$
  is broadcasted to the Processes $( p_1 + \tilde{p} \chi )$ $( \tilde{p} = 0, 1, ..., \chi - 1 )$ except the Process $( p_1 + p_1 \chi )$ itself as $\check{T}_{n_{d-1} \dots n_1 n_{d-1}^{\prime} \dots n_1^{\prime} c}^{c^{\prime} = \rho}$.
  After this broadcasting, contractions in (\ref{def_hat_S_N}) and (\ref{def_hat_S_P}) are done without communication between processes.
  Elements of the tensor $\hat{S}_{\hat{f} \hat{g} \rho \eta}^{\prime ( j )}$ are gathered to the Process $( j - 1 )$ $( j = 1, ..., d - 1 )$.
  Similarly, those of the tensor $\hat{S}_{\hat{f} \hat{g} \rho \eta}^{( j )}$ are gathered to the Process $( d - 2 + j )$ $( j = 1, ..., d - 1 )$.

\begin{figure}[H]
      \begin{center}
         \includegraphics[width=0.9\textwidth]{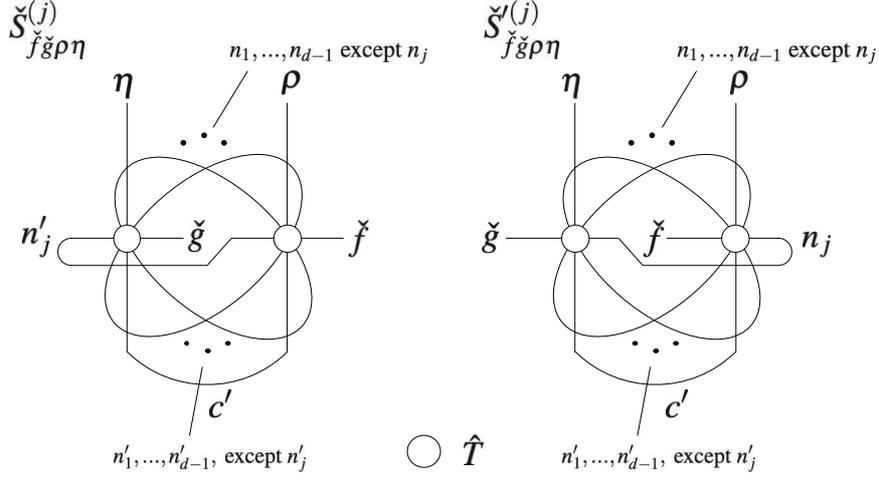}
         \caption{Representation of tensors $\check{S}_{\check{f} \check{g} \rho \eta}^{( j )}$ and $\check{S}_{\check{f} \check{g} \rho \eta}^{\prime ( j )}$ by tensor network}
         \label{Fig_TensorsS_check}
      \end{center}
  \end{figure}

   In the second step, for $j = 1, ..., d - 1$, the following tensors are computed from two $c$-fixed tensors.
  \begin{gather}
      \check{S}_{\check{f} \check{g} \rho \eta}^{( j )}
      = \sum \hat{T}_{n_{d-1} \dots n_{j + 1} \check{f}     n_{j - 1} \dots n_1 n_{d-1}^{\prime} \dots n_1^{\prime} c^{\prime} }^{c= \rho}
                  \hat{T}_{n_{d-1} \dots n_{j + 1} \check{g} n_{j - 1} \dots n_1 n_{d-1}^{\prime} \dots n_1^{\prime} c^{\prime} }^{c= \eta},    \label{def_check_S_N}  \\
      \check{S}_{\check{f} \check{g} \rho \eta}^{\prime ( j )}
      = \sum \nolimits' \hat{T}_{n_{d-1} \dots n_1 n_{d-1}^{\prime} \dots n_{j + 1}^{\prime} \check{f} n_{j - 1}^{\prime} \dots n_1^{\prime} c^{\prime} }^{c = \rho}
                                 \hat{T}_{n_{d-1} \dots n_1 n_{d-1}^{\prime} \dots n_{j + 1}^{\prime} \check{g}  n_{j - 1}^{\prime} \dots n_1^{\prime} c^{\prime} }^{c = \eta},  \label{def_check_S_P}
  \end{gather}
  where $\sum$ and $\sum^{\prime}$ are
  \begin{equation}  \label{contraction_check_S_N}
      \sum = \sum_{n_{d - 1} = 0}^{\chi_{n_{d - 1}} - 1} \cdots \sum_{n_{j + 1} = 0}^{\chi_{n_{j + 1} } - 1}
                  \sum_{n_{j - 1} = 0}^{\chi_{n_{j - 1}} - 1} \cdots \sum_{n_1 = 0}^{\chi_{n_1} - 1}
                  \sum_{n_{d - 1}^{\prime} = 0}^{\chi_{n_{d - 1}} - 1} \cdots \sum_{n_1^{\prime} = 0}^{\chi_{n_1} - 1}
                  \sum_{c^{\prime} = 0}^{\chi_c - 1}
  \end{equation}
  and
  \begin{equation} \label{contraction_check_S_P}
      \sum \nolimits' = \sum_{n_{d - 1} = 0}^{\chi_{n_{d - 1}} - 1} \cdots \sum_{n_1 = 0}^{\chi_{n_1} - 1}
                                \sum_{n_{d - 1}^{\prime} = 0}^{\chi_{n_{d - 1}} - 1} \cdots \sum_{n_{j + 1}^{\prime} = 0}^{\chi_{n_{j + 1} } - 1}
                                \sum_{n_{j - 1}^{\prime} = 0}^{\chi_{n_{j - 1}} - 1} \cdots \sum_{n_1^{\prime} = 0}^{\chi_{n_1} - 1}
                                \sum_{c^{\prime} = 0}^{\chi_c - 1},
  \end{equation}
  respectively.
  Their representations by tensor network is shown in Fig. \ref{Fig_TensorsS_check}.
  These representations correspond to lower half of tensor network shown in Fig. \ref{FIg_Matrix_SVD}.
  In our method, The elements of $\check{S}_{\check{f} \check{g} \rho \eta}^{( j )}$ and $\check{S}_{\check{f} \check{g} \rho \eta}^{\prime ( j )}$ are computed in the Process $\rho + \eta \chi$.
  The $c$-fixed tensor $\hat{T}_{n_{d-1} \dots n_1 n_{d-1}^{\prime} \dots n_1^{\prime} c^{\prime}}^{c = \rho}$ have already been stored in each process by broadcasting explained in Section \ref{BroadcastBeginning}.
  The $c$-fixed tensor in the Process $( p_2 + p_2 \chi )$
  is broadcasted to the Processes $( \tilde{p} + p_2 \chi )$ $( \tilde{p} = 0, 1, ..., \chi - 1 )$ except the Process $( p_2 + p_2 \chi )$ itself as $\hat{T}_{n_{d-1} \dots n_1 n_{d-1}^{\prime} \dots n_1^{\prime} c^{\prime}}^{c = \eta}$.
  After this broadcasting, contractions (\ref{def_check_S_N}) and (\ref{def_check_S_P}) are done without communication between processes.
  Elements of the tensor $\check{S}_{\check{f} \check{g} \rho \eta}^{\prime ( j )}$ are gathered to the Process $( j - 1 )$ $( j = 1, ..., d - 1 )$.
  Similarly, those of the tensor $\check{S}_{\check{f} \check{g} \rho \eta}^{( j )}$ are gathered to the Process $( d - 2 + j )$ $( j = 1, ..., d - 1 )$.

  In the last step, contractions
  \begin{equation} 
      A_{( \hat{f} \check{f} ), ( \hat{g} \check{g} )}^{( j )}
      =\sum_{\rho = 0}^{\chi_c - 1} \sum_{\eta = 0}^{\chi_c - 1} \hat{S}_{\hat{f} \hat{g} \rho \eta}^{( j )} \check{S}_{\check{f} \check{g} \rho \eta}^{( j )},
  \end{equation}
  are done in the Processes $( j - 1 )$ $( j = 1, ..., d - 1 )$ and those
  \begin{equation} 
      A_{( \hat{f} \check{f} ), ( \hat{g} \check{g} )}^{\prime ( j )}
      =\sum_{\rho = 0}^{\chi_c - 1} \sum_{\eta = 0}^{\chi_c - 1} \hat{S}_{\hat{f} \hat{g} \rho \eta}^{\prime ( j )} \check{S}_{\check{f} \check{g} \rho \eta}^{\prime ( j )},
  \end{equation}
  are done in the Processes $( d - 2 + j )$ $( j = 1, ..., d - 1 )$.
  Thus, matrices to which SVD is applied are obtained.

  \subsubsection{Singular value decomposition}  \label{SingularValueDecomposition}

  In the Processes $( j - 1 )$ $( j = 1, ..., d - 1 )$, singular value decomposition
  \begin{equation}   \label{SVD_N}
      A^{( j )} = U^{( j )} \Sigma^{( j )} ( V^{( j )} )^{\top}
  \end{equation}
  is executed.
  In the Processes $( d - 2 + j  )$ $( j = 1, ..., d - 1 )$, singular value decomposition
  \begin{equation}   \label{SVD_P}
      A^{\prime ( j )} = U^{\prime ( j )} \Sigma^{\prime ( j )} ( V^{\prime ( j )} )^{\top}
  \end{equation}
  is executed.
  Singular values in the matrices $\Sigma^{( j )}$ and $\Sigma^{\prime ( j )}$ are ordered in descending order,
  namely, $\sigma_1^{( j )} \geq \sigma_2^{( j )} \geq \cdots \geq \sigma_{\chi_j^2}^{( j )}$ and $\sigma_1^{\prime ( j )} \geq \sigma_2^{\prime ( j )} \geq \cdots \geq \sigma_{\chi_j^2}^{\prime ( j )}$, respectively.

  \subsubsection{Computation of judgment values for choice of unitary matrices}  \label{Judgment_Values}

  For $j = 1, ..., d - 1$, one of the unitary matrices $U^{( j )}$ in (\ref{SVD_N}) and $U^{\prime ( j )}$ in (\ref{SVD_P}) is chosen for construction of a new local tensor used in the next coarse-graining procedure.
  Such unitary matrices are chosen according to some criterion.
  This criterion is not unique.
  Any criterion is acceptable as long as it is rational.
  For example, we can adopt a criterion given in \cite{XCQZYX12} which is reviewed in Section \ref{HOTRG}.
  To choose one of the unitary matrices, judgment values $\varepsilon^{( j )}$ and $\varepsilon^{\prime ( j )}$ for $U^{( j )}$ and $U^{\prime ( j )}$, respectively, are computed.

  \subsubsection{Choice of unitary matrices}  \label{ChoiceUnitaryMatrices}

  We choose unitary matrices.
  For $j = 1, ..., d - 1$, unitary matrix $U^{( j )}$ and judgment value $\varepsilon^{( j )}$ are stored in the Processes $( d - 2 + j )$,
                                    and unitary matrix $U^{\prime ( j )}$ and judgment value $\varepsilon^{\prime ( j )}$ are stored in the Processes $( j - 1 )$.
  Then, the unitary matrix $U^{( j )}$ and the judgment value $\varepsilon^{( j )}$ in the Process $( d - 2 + j )$ are transferred to the Process $( j - 1 )$ for comparison.
  For $j = 1, ..., d - 1$, one of the two unitary matrices $U^{( j )}$ and $U^{\prime ( j )}$ is chosen for construction of a new local tensor for the next coarse-graining procedure according to some criterion.

  \subsubsection{Truncation and distribution of the chosen unitary matrices}  \label{TruncationDistribution}
  
  For $j = 1, ..., d - 1$, let us denote a unitary matrix chosen among $U^{( j )}$ and $U^{\prime ( j )}$ by $\tilde{U}^{( j )}$.
  The first $\chi_j^{(\textrm{new})}$ columns of the chosen unitary matrix $\tilde{U}^{( j )}$ are used to construct a new local tensor for the next coarse-graining procedure.
  For each column vector, we may invert it. 
  The column vectors are gathered to the Process 0.
  Then, they are broadcasted from the Process 0 to the other processes.

  \subsubsection{Contractions for a new local tensor}  \label{ContractionsForNewTensor}
  
  We construct a new local tensor from two local tensors and $2 ( d - 1 )$ unitary matrices.
  Equation (\ref{Contraction_NewTensor}) on new local tensor is rewritten by using the $c$-fixed and the $c^{\prime}$-fixed tensors as
  \begin{align}
      T_{n_{d - 1} \dots n_1 n_{d - 1}^{\prime} \dots n_1^{\prime} c_0 c_0^{\prime}}^{(\textrm{new})}
      =\sum &\left( \tilde{U}_{( \hat{n}_{d - 1}^{\prime} \check{n}_{d - 1}^{\prime} ), n_{d - 1}^{\prime}}^{( d - 1 )} \cdots \tilde{U}_{( \hat{n}_1^{\prime} \check{n}_1^{\prime} ), n_1^{\prime}}^{( 1 )} 
                            \check{T}_{\check{n}_{d - 1} \dots \check{n}_1 \check{n}_{d - 1}^{\prime} \dots \check{n}_1^{\prime} \rho}^{c^{\prime} = c_0^{\prime}} \right.  \notag \\
                 &~~~ \times \left. \hat{T}_{\hat{n}_{d - 1} \dots \hat{n}_1 \hat{n}_{d - 1}^{\prime} \dots \hat{n}_1^{\prime} \rho}^{c = c_0}
                                      \tilde{U}_{( \hat{n}_1 \check{n}_1 ), n_1}^{( 1 )} \cdots \tilde{U}_{( \hat{n}_{d - 1} \check{n}_{d - 1} ), n_{d - 1}}^{( d - 1 )} \right) ,
  \end{align}
  where $\sum$ is
  \begin{equation}
     \sum_{\check{n}_{d - 1}^{\prime} = 0}^{\chi_{n_{d - 1}} - 1}
     \cdots
     \sum_{\check{n}_1^{\prime} = 0}^{\chi_{n_1} - 1}
     \sum_{\hat{n}_{d - 1}^{\prime} = 0}^{\chi_{n_{d - 1}} - 1}
     \cdots
     \sum_{\hat{n}_1^{\prime} = 0}^{\chi_{n_1} - 1}
     \sum_{\check{n}_{d - 1} = 0}^{\chi_{n_{d - 1}} - 1}
     \cdots
     \sum_{\check{n}_1 = 0} ^{\chi_{n_1} - 1}
     \sum_{\hat{n}_{d - 1} = 0}^{\chi_{n_{d - 1}} - 1}
     \cdots
     \sum_{\hat{n}_1 = 0}^{\chi_{n_1} - 1}
     \sum_{\rho = 0}^{\chi_c - 1}.
  \end{equation}
  For fixed $c_0$ and $c_0^{\prime}$, elements of $T_{n_{d - 1} \dots n_1 n_{d - 1}^{\prime} \dots n_1^{\prime} c_0 c_0^{\prime}}^{(\textrm{new})}$ are computed in the Process $( c_0 + c_0^{\prime} \chi )$.
  The elements of the $c$-fixed and the $c^{\prime}$-fixed tensors have already been stored by broadcasting described in Section \ref{BroadcastBeginning}.
  No communication between processes occurs during this contraction.
  In our method, this contraction is done for fixed $n_1^{\prime}, ..., n_{d - 1}^{\prime}$ to avoid increase of memory space requirement.
  Contraction procedure for fixed $n_1^{\prime}, ..., n_{d - 1}^{\prime}$ consists of the following three steps.

\begin{figure}[H]
      \begin{center}
         \includegraphics[width=0.4\textwidth]{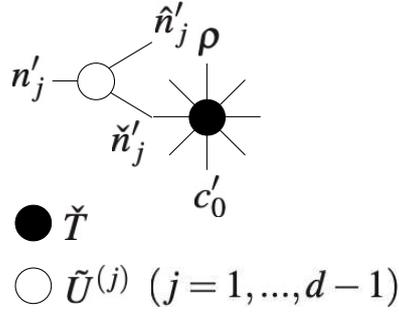}
         \caption{Tensor network representation of Eqs. from (\ref{TU1}) to (\ref{TUNCD})}
         \label{Fig_ContractEarly}
      \end{center}
  \end{figure}

  In the first step, contraction among the $c^{\prime}$-fixed tensor and the unitary matrices in the Prime side is done.
  Namely, we compute
  \begin{gather}
      \check{C}_{\check{n}_{d - 1} \dots \check{n}_1 \check{n}_{d - 1}^{\prime} \dots \check{n}_2^{\prime} \hat{n}_1^{\prime} \rho}^{( 1 ), c^{\prime} = c_0^{\prime}, n_1^{\prime}}
      = \sum_{\check{n}_1^{\prime} = 0}^{\chi_1 - 1} \tilde{U}_{( \hat{n}_1^{\prime} \check{n}_1^{\prime} ), n_1^{\prime}}^{( 1 )} 
                                                                                \check{T}_{\check{n}_{d - 1} \dots \check{n}_1 \check{n}_{d - 1}^{\prime} \dots \check{n}_1^{\prime} \rho}^{c^{\prime} = c_0^{\prime}},  \label{TU1} \\
      \check{C}_{\check{n}_{d - 1} \dots \check{n}_1 \check{n}_{d - 1}^{\prime} \dots \check{n}_3^{\prime} \hat{n}_2^{\prime} \hat{n}_1^{\prime} \rho}^{( 2 ), c^{\prime} = c_0^{\prime}, n_2^{\prime} n_1^{\prime}}
      = \sum_{\check{n}_2^{\prime} = 0}^{\chi_2 - 1} \tilde{U}_{( \hat{n}_2^{\prime} \check{n}_2^{\prime} ), n_2^{\prime}}^{( 2 )} 
                                                                                \check{C}_{\check{n}_{d - 1} \dots \check{n}_1 \check{n}_{d - 1}^{\prime} \dots \check{n}_2^{\prime} \hat{n}_1^{\prime} \rho}^{( 1 ), c^{\prime} = c_0^{\prime}, n_1^{\prime}},  \\
      \qquad  \vdots \notag  \\
      \check{C}_{\check{n}_{d - 1} \dots \check{n}_1 \hat{n}_{d - 1}^{\prime} \dots \hat{n}_1^{\prime} \rho}^{( d - 1 ), c^{\prime} = c_0^{\prime}, n_{d - 1}^{\prime} \dots n_1^{\prime}}
      = \sum_{\check{n}_{d - 1}^{\prime} = 0}^{\chi_{d - 1} - 1} 
         \tilde{U}_{( \hat{n}_{d - 1}^{\prime} \check{n}_{d - 1}^{\prime} ), n_{d - 1}^{\prime}}^{( d - 1 )} 
         \check{C}_{\check{n}_{d - 1} \dots \check{n}_1 \check{n}_{d - 1}^{\prime} \hat{n}_{d - 2}^{\prime} \dots \hat{n}_1^{\prime} \rho}^{( d - 2 ), c^{\prime} = c_0^{\prime}, n_{d - 2}^{\prime} \dots n_1^{\prime}}.  \label{TUNCD}
  \end{gather}
  See Fig. \ref{Fig_ContractEarly} for help of understanding.

\begin{figure}[H]
      \begin{center}
         \includegraphics[width=0.4\textwidth]{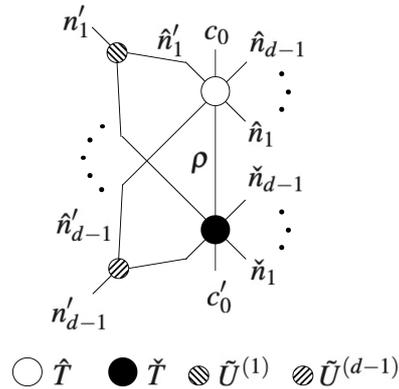}
         \caption{Tensor network representation of Eq. (\ref{BottleneckPart})}
         \label{Fig_ContractBN}
      \end{center}
  \end{figure}

  In the second step, contraction among the tensor $\check{C}_{\check{n}_{d - 1} \dots \check{n}_1 \hat{n}_{d - 1}^{\prime} \dots \hat{n}_1^{\prime} \rho}^{( d - 1 ), c^{\prime} = c_0^{\prime}, n_{d - 1}^{\prime} \dots n_1^{\prime}}$ and the $c$-fixed tensor is done.
  By this contraction, we have
  \begin{equation}
      \hat{C}_{\check{n}_{d - 1} \dots \check{n}_1 \hat{n}_{d - 1} \dots \hat{n}_1}^{( d - 1 ), ( c, c^{\prime} ) = ( c_0, c_0^{\prime} ), n_{d - 1}^{\prime} \dots n_1^{\prime}}
      = \sum \hat{T}_{\hat{n}_{d - 1} \dots \hat{n}_1 \hat{n}_{d - 1}^{\prime} \dots \hat{n}_1^{\prime} \rho}^{c = c_0}
                  \check{C}_{\check{n}_{d - 1} \dots \check{n}_1 \hat{n}_{d - 1}^{\prime} \dots \hat{n}_1^{\prime} \rho}^{( d - 1 ), c^{\prime} = c_0^{\prime}, n_{d - 1}^{\prime} \dots n_1^{\prime}},  \label{BottleneckPart}
  \end{equation}
  where $\sum$ is
  \begin{equation}
     \sum_{\hat{n}_{d - 1}^{\prime} = 0}^{\chi_{n_{d - 1}} - 1}
     \cdots
     \sum_{\hat{n}_1^{\prime} = 0}^{\chi_{n_1} - 1}
     \sum_{\rho = 0}^{\chi_c - 1}.
  \end{equation}
  See Fig. \ref{Fig_ContractBN} for help of understanding.

\begin{figure}[H]
      \begin{center}
         \includegraphics[width=0.3\textwidth]{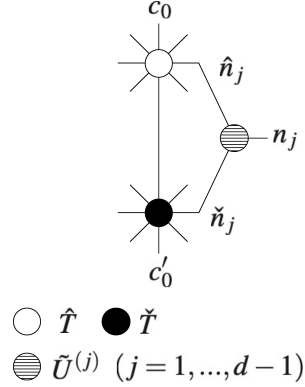}
         \caption{Tensor network representation of Eqs. from (\ref{AfterBNNCD}) to (\ref{AfterBN1})}
         \label{Fig_ContractLate}
      \end{center}
  \end{figure}

  In the last step, contraction among the tensor $\hat{C}_{\check{n}_{d - 1} \dots \check{n}_1 \hat{n}_{d - 1} \dots \hat{n}_1}^{( d - 1 ), ( c, c^{\prime} ) = ( c_0, c_0^{\prime} ), n_{d - 1}^{\prime} \dots n_1^{\prime}}$
                                                                and the unitary matrices in the Non-prime side is done.
  Namely, we compute
  \begin{gather}
      \hat{C}_{n_{d - 1} \check{n}_{d - 2} \dots \check{n}_1 \hat{n}_{d - 2} \dots \hat{n}_1}^{( d - 2 ), ( c, c^{\prime} ) = ( c_0, c_0^{\prime} ), n_{d - 1}^{\prime} \dots n_1^{\prime}}  \notag  \\
      = \sum_{\hat{n}_{d - 1} = 0}^{\chi_{d - 1} - 1} \sum_{\check{n}_{d - 1} = 0}^{\chi_{d - 1} - 1}
         \tilde{U}_{( \hat{n}_{d - 1} \check{n}_{d - 1} ), n_{d - 1}}^{( d - 1 )} 
         \hat{C}_{\check{n}_{d - 1} \dots \check{n}_1 \hat{n}_{d - 1} \dots \hat{n}_1}^{( d - 1 ), ( c, c^{\prime} ) = ( c_0, c_0^{\prime} ), n_{d - 1}^{\prime} \dots n_1^{\prime}},  \label{AfterBNNCD}  \\
      \qquad  \vdots \notag  \\
      \hat{C}_{n_{d - 1} \dots n_2 \check{n}_1 \hat{n}_1}^{( 1 ), ( c, c^{\prime} ) = ( c_0, c_0^{\prime} ), n_{d - 1}^{\prime} \dots n_1^{\prime}}
      = \sum_{\hat{n}_2 = 0}^{\chi_2- 1} \sum_{\check{n}_2 = 0}^{\chi_2- 1} \tilde{U}_{( \hat{n}_2 \check{n}_2 ), n_2}^{( 2 )}
                                                                                                                     \hat{C}_{n_{d - 1} \dots n_3 \check{n}_2 \check{n}_1 \hat{n}_2 \hat{n}_1}^{( 2 ), ( c, c^{\prime} ) = ( c_0, c_0^{\prime} ), n_{d - 1}^{\prime} \dots n_1^{\prime}},  \\
      \hat{C}_{n_{d - 1} \dots n_1}^{( 0 ), ( c, c^{\prime} ) = ( c_0, c_0^{\prime} ), n_{d - 1}^{\prime} \dots n_1^{\prime}}
      = \sum_{\hat{n}_1 = 0}^{\chi_1 - 1} \sum_{\check{n}_1 = 0}^{\chi_1 - 1} \tilde{U}_{( \hat{n}_1 \check{n}_1 ), n_1}^{( 1 )} 
                                                                                                                       \hat{C}_{n_{d - 1} \dots n_2 \check{n}_1 \hat{n}_1}^{( 1 ), ( c, c^{\prime} ) = ( c_0, c_0^{\prime} ), n_{d - 1}^{\prime} \dots n_1^{\prime}}. \label{AfterBN1}
  \end{gather}
  See Fig. \ref{Fig_ContractLate} for help of understanding.
  Elements of a new local tensor is given as
  \begin{equation}  \label{Elements_NewTensor}
      T_{n_{d - 1} \dots n_1 n_{d - 1}^{\prime} \dots n_1^{\prime} c_0 c_0^{\prime}}^{(\textrm{new})}
      = \hat{C}_{n_{d - 1} \dots n_1}^{( 0 ), ( c, c^{\prime} ) = ( c_0, c_0^{\prime} ), n_{d - 1}^{\prime} \dots n_1^{\prime}}.
  \end{equation}
  
  In implementation of our method, these contractions are executed using loops for $n_1^{\prime}$, ..., $n_{d - 1}^{\prime}$.
  The loop for the index $n_1^{\prime}$ is the outermost one and that for the index $n_{d - 1}^{\prime}$ is the innermost one.
  Then, for quantities which appear in this contraction procedure, indices as superscripts are identified through counters of loops or a process number.
  Those as subscripts are identified through an element number of an array.
  Thus, memory space requirement in each process in this contraction procedure is kept to be $O ( \chi^{2d - 1} )$.

  In the case of $d \geq 3$, the bottleneck part of the HOTRG in computational cost is the above-mentioned second step.
  Thus computational cost of our method in each process is $O ( \chi^{4d - 3} )$ when the dimensionality $d$ of a lattice is $d \geq 3$.


  \subsubsection{Preparation for the new $c$-fixed tensor, the trace of the new tensor and the ratio used to identify a phase of a model}  \label{PreparationNonprime}

  Now, we have obtained all elements of the new tensor.
  As mentioned in Section \ref{Directions_TN}, directions are renamed for the next coarse-graining procedure.
  The direction Next 1 in the present coarse-graining procedure is the direction Current in the next coarse-graining procedure.
  Then, as preparation to construct the state at the beginning of the next coarse-graining procedure on the $c$-fixed tensor (See Section \ref{StateBeginning}.),
  we transfer elements $T_{n_{d - 1} \dots n_1 n_{d - 1}^{\prime} \dots n_1^{\prime} c c^{\prime}}^{(\textrm{new})}$ of a new local tensor to an appropriate process.
  Among these elements,
  those of which the index $n_1$ is $\tilde{n}$ $( \tilde{n} = 0, 1, ..., \chi_1^{(\textrm{new})} - 1 )$, namely,
  the elements $T_{n_{d - 1} \dots n_2 \tilde{n} n_{d - 1}^{\prime} \dots n_2^{\prime} n_1^{\prime} c c^{\prime}}^{(\textrm{new})}$ are gathered to the Process $( \tilde{n} + \tilde{n} \chi )$.
  This is also preparation to compute the trace of the new tensor and the ratio used to identify a phase of a model.

  \subsubsection{Computation of the trace of the new local tensor and the ratio used to identify a phase of a model}  \label{ComputeTrRatio}

  We compute the trace of the new local tensor and the ratio $X$ in (\ref{ratio_X}).
  The matrix $Y_{c, c^{\prime}} $ in (\ref{Matrix_Y}) is given as
  \begin{equation}
      Y_{c, c^{\prime}} =  \sum_{\tilde{n}_{d - 1} = 0}^{\chi_{d - 1}^{(\textrm{new})} - 1}  \cdots \sum_{\tilde{n}_1 = 0}^{\chi_1^{(\textrm{new})} - 1} T_{\tilde{n}_{d-1} \dots \tilde{n}_1 \tilde{n}_{d-1} \dots \tilde{n}_1 c c^{\prime}}^{(\textrm{new})}.
  \end{equation}
  Obviously, $\textrm{trace} Y$ is equal to $\textrm{Tr} T^{(\textrm{new})}$.
  Let us introduce matrices $Q_{c, c^{\prime}}^{( p_1, p_2 )}$ $( p_1, p_2 = 0, 1, ..., \chi - 1 )$.
  For particular $p_1$ and $p_2$, the matrix $Q_{c, c^{\prime}}^{( p_1, p_2 )}$ is stored in the Process $( p_1 + p_2 \chi )$.
  In the Processes $( \tilde{n} + \tilde{n} \chi )$ $( \tilde{n} = 0, 1, ..., \chi_1^{(\textrm{new})} - 1 )$, the matrix $Q_{c, c^{\prime}}^{( \tilde{n} , \tilde{n}  )}$ is computed as
   \begin{equation}
      Q_{c, c^{\prime}}^{( \tilde{n} , \tilde{n}  )}
      = \sum_{\tilde{n}_{d - 1} = 0}^{\chi_{d - 1}^{(\textrm{new})} - 1} \cdots \sum_{\tilde{n}_2 = 0}^{\chi_2^{(\textrm{new})} - 1} T_{\tilde{n}_{d-1} \dots \tilde{n}_2 \tilde{n} \tilde{n}_{d-1} \dots \tilde{n}_2 \tilde{n} c c^{\prime}}^{(\textrm{new})}
  \end{equation}
  using the gathered elements explained in Section \ref{PreparationNonprime}.
  In the other processes, this matrix is set to zero matrix.
  Then we compute
   \begin{equation}  \label{ALLREDUCE_Y}
      Y_{c, c^{\prime}} =  \sum_{p_1= 0}^{\chi - 1} \sum_{p_2= 0}^{\chi - 1} Q_{c, c^{\prime}}^{( p_1, p_2 )}
  \end{equation}
  and the result are stored in all the processes.
  In this step, communication between processes occurs.
  After computation of $Y_{c, c^{\prime}}$, computation of the trace and the ratio $X$ is straightforwardly done in each process without communication between processes.

  \subsubsection{Construction of the new $c$-fixed tensor before normalization}  \label{ConstructNewFixT_c_N}

  In the Processes $( \tilde{n} + \tilde{n} \chi )$ $( \tilde{n} = 0, 1, ..., \chi_1^{(\textrm{new})} - 1 )$, 
  indices of the elements  $T_{n_{d - 1} \dots n_2 \tilde{n} n_{d - 1}^{\prime} \dots n_2^{\prime} n_1^{\prime} c c^{\prime}}^{(\textrm{new})}$ shown in Section \ref{PreparationNonprime} should be renamed to be suitable for the next coarse-graining procedure.
  Then, the new $c$-fixed tensor before normalization is constructed as
   \begin{equation}
      \hat{T}_{\bar{n}_{d-1} \dots \bar{n}_1 \bar{n}_{d-1}^{\prime} \dots \bar{n}_1^{\prime}  \bar{c}^{\prime}}^{c = \tilde{n}, \textrm{Not normalized}}
      = T_{n_{d - 1} \dots n_2 \tilde{n} n_{d - 1}^{\prime} \dots n_2^{\prime} n_1^{\prime} c c^{\prime}}^{(\textrm{new})},
  \end{equation}
  where
  \begin{gather}
      \bar{c}^{\prime} = n_1^{\prime},  \\
      \bar{n}_j = n_{j + 1},  \qquad  (j = 1, ..., d - 2),   \\
      \bar{n}_j^{\prime} = n_{j + 1}^{\prime},  \qquad  (j = 1, ..., d - 2),   \\
      \bar{n}_{d - 1} = c,   \\
      \bar{n}_{d - 1}^{\prime} = c^{\prime}.
  \end{gather}

  \subsubsection{Construction of the new $c^{\prime}$-fixed tensor before normalization}  \label{ConstructNewFixT_c_P}

  The new $c^{\prime}$-fixed tensor before normalization is constructed in a way similar to Sections \ref{PreparationNonprime} and \ref{ConstructNewFixT_c_N}.
  Among elements $T_{n_{d - 1} \dots n_1 n_{d - 1}^{\prime} \dots n_1^{\prime} c c^{\prime}}^{(\textrm{new})}$ of a new local tensor,
  those of which the index $n_1^{\prime}$ is $\tilde{n}$ $( \tilde{n} = 0, 1, ..., \chi_1^{(\textrm{new})} - 1 )$, namely,
  the elements $T_{n_{d - 1} \dots n_2 n_1 n_{d - 1}^{\prime} \dots n_2^{\prime} \tilde{n} c c^{\prime}}^{(\textrm{new})}$ are gathered to the Process $( \tilde{n} + \tilde{n} \chi )$.
  We construct the new $c^{\prime} $-fixed tensor as
   \begin{equation}
      \check{T}_{\bar{n}_{d-1} \dots \bar{n}_1 \bar{n}_{d-1}^{\prime} \dots \bar{n}_1^{\prime}  \bar{c}}^{c^{\prime} = \tilde{n}, \textrm{Not normalized}}
      = T_{n_{d - 1} \dots n_2 n_1 n_{d - 1}^{\prime} \dots n_2^{\prime} \tilde{n} c c^{\prime}}^{(\textrm{new})},
  \end{equation}
  where
  \begin{gather}
      \bar{c} = n_1,   \\
      \bar{n}_j = n_{j + 1},  \qquad  (j = 1, ..., d - 2),   \\
      \bar{n}_j^{\prime} = n_{j + 1}^{\prime},  \qquad  (j = 1, ..., d - 2),   \\
      \bar{n}_{d - 1} = c,   \\
      \bar{n}_{d - 1}^{\prime} = c^{\prime}.
  \end{gather}

  \subsubsection{Bond dimensions in the next coarse-graining procedure}  \label{BondDimensions_Next}

  Bond dimensions in the next coarse-graining procedure are set.
  Let us denote bond dimensions of the new local tensor in the directions Current and Next $j$ $( j = 1, ..., d - 1 )$ in the next coarse-graining procedure by $\bar{\chi}_c$ and $\bar{\chi}_{n_j}$, respectively.
  As shown in Section \ref{LocalTensors}, considering renaming of directions, we should set $\bar{\chi}_c$ and $\bar{\chi}_{n_j}$ $( j = 1, ..., d - 1 )$ as
  \begin{gather}
      \bar{\chi}_c = \chi_{n_1}^{(\textrm{new})}, \\
      \bar{\chi}_{n_j} = \chi_{n_{j + 1}}^{(\textrm{new})}, \qquad  ( j = 1, ..., d - 2 ),  \\
      \bar{\chi}_{n_{d - 1}} = \chi_c.
  \end{gather}

  \subsubsection{Computation of normalization factor}  \label{ComputeNormalizationF}

  A normalization factor is computed.
  When we use the inverse of the trace of a local tensor in (\ref{Trace_LocalTensor}) as a normalization factor, the trace has already been obtained in Section \ref{ComputeTrRatio} as the trace of the matrix $Y_{c, c^{\prime}}$.

  \subsubsection{Computation of physical quantities}  \label{ComputePhysics}

  Assume that we finish the $p$-th $( p =1, 2, ... )$ coarse-graining procedure.

  Using (\ref{eq_lnZ}), we have the quantity $\ln Z^{( p )}$.
  The trace $\textrm{Tr} T_1^{(p)}$ in (\ref{eq_lnZ}) has already been obtained in Section \ref{ComputeTrRatio} as the trace of the matrix $Y_{c, c^{\prime}}$.
  The quantity $K^{( p )}$ in (\ref{eq_lnZ}) has already been obtained after the $( p - 1 )$-th coarse-graining procedure.
  We also obtain the quantity $\ln Z^{( p )} / V^{( p )}$, where $V^{( p )} = 2^p$, by straightforward computation.
  
  Let the normalization factor obtained in Section \ref{ComputeNormalizationF} be denoted by $f^{( p )}$.
  Substituting it into (\ref{K_recurrence}), we have the quantity $K^{( p + 1 )}$.

  \subsubsection{Normalization of the new $c$-fixed and $c^{\prime}$-fixed tensors}  \label{Normalization_NewTensors}

  All the elements of  the new $c$-fixed and $c^{\prime}$-fixed tensors obtained in Sections \ref{ConstructNewFixT_c_N} and \ref{ConstructNewFixT_c_P}, respectively,
  are multiplied by normalization factor obtained in Section \ref{ComputeNormalizationF}.
  Then we have $c$-fixed and $c^{\prime}$-fixed tensors for the next coarse-graining procedure as
  \begin{gather}
      \hat{T}_{n_{d-1} \dots n_1n_{d-1}^{\prime} \dots n_1^{\prime} c^{\prime}}^{c = c_0}
      = f^{( p )} \hat{T}_{n_{d-1} \dots n_1n_{d-1}^{\prime} \dots n_1^{\prime} c^{\prime}}^{c = c_0, \textrm{Not normalized}},  \\
      \check{T}_{n_{d-1} \dots n_1 n_{d-1}^{\prime} \dots n_1^{\prime} c}^{c^{\prime} = c_0^{\prime}}
      = f^{( p )} \check{T}_{n_{d-1} \dots n_1 n_{d-1}^{\prime} \dots n_1^{\prime} c}^{c^{\prime} = c_0^{\prime}, \textrm{Not normalized}}.
  \end{gather}

 \section{Numerical experiments}  \label{NumExp}

  In this section, we execute numerical experiments.
  We consider the well-known Ising model.
  The experiment is executed to discuss computational cost.
  In this experiment, we consider a four-dimensional $( d = 4 )$ simple lattice and measure elapsed time.

  We execute these experiments on Oakforest-PACS.
  Source codes are compiled by using command mpiifort.
  For parallel computing, compile options -parallel , -qopenmp  and -mkl=parallel are used.
  For optimization, compile options -axMIC-AVX512 and -O3 are used.
  Computation is executed on $\chi^2$ nodes for a specified bond dimension $\chi$.
  On each node, one process of MPI runs.
  The number of threads in OpenMP is 64.

\begin{figure}[H]
      \begin{center}
         \includegraphics[width=0.6\textwidth]{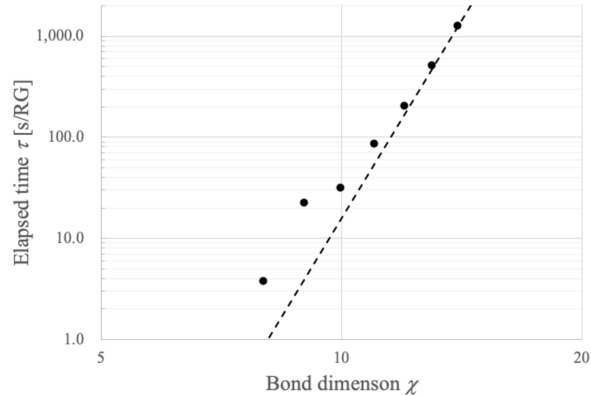}
         \caption{Elapsed times of the presented method}
         \label{Fig_Exp_4D}
      \end{center}
  \end{figure}

  In the experiment, we measure elapsed time for once coarse-graining procedure varying bond dimension $\chi$ from 8 to 14.
  At the beginning of computation, bond dimensions are $2$ in all the directions.
  Bond dimensions are updated as explained in Section \ref{LocalTensors}.
  Then, elapsed time of  the first, the second and the third coarse-graining procedures cannot be adopted as data.
  Let us regard four times of sequential coarse-graining procedures as a set since we consider a four-dimensional simple lattice.
  Then, from the first to the fourth coarse-graining procedures belong to the first set.
  Thus, we adopt elapsed time form the fifth to the twenty-fourth coarse-graining procedures (from the second to the sixth sets) as data.
  Averages of elapsed times of twenty times of coarse-graining procedures are plotted in Fig. \ref{Fig_Exp_4D}.
  Vertical axis represents elapsed time $\tau$ for once coarse-graining procedure in second and horizontal axis represents bond dimension.
  They are in logarithmic scale.
  The word RG in the label of the vertical axis is abbreviation of Renormalization Group.
  Since computational cost in each process is $O (\chi^{4d - 3}) = O (\chi^{13})$, a dashed line which represents a relationship
  \begin{equation}
      \tau = \alpha \chi^{13},
  \end{equation}
  where $\alpha$ is determined to make this line passes the plotted point of $\chi = 14$, is added to this graph.
  The plotted points seem to approach this line asymptotically.
  Elapsed times for once coarse-graining procedure are 510.76 and 1247.65 in second for bond dimensions 13 and 14, respectively.

 \section{Concluding remarks}  \label{ConcludingRmks}
  
  A parallel computing algorithm for the Higher Order Tensor Renormalization Group is presented.
  Computational cost $O ( \chi^{4d - 1} )$ and memory space requirement $O ( \chi^{2d} )$ of the HOTRG in a $d$-dimensional simple lattice model is not cheap when we consider higher dimensional model.
  When we distribute elements of a local tensor to each process in the simplest way such that an element is placed to one process, and execute the HOTRG in parallel, we would be suffered from cost for communication between processes.
  This problem in cost for communication is caused since we have to get elements of a local tensor which are necessary for contraction procedure from another process.
  In our method, we place sufficient local tensor elements for a considering contraction step to avoid communication between processes and accept placement of an element to more than one process.
  Distribution of elements of local tensors are determined by one of indices of each local tensor which are not contracted during the considering contraction procedure.
  In the cases of $d \geq 3$, computational cost in each process is $O ( \chi^{4d - 3} )$ and memory space requirement in each process is $O ( \chi^{2d - 1} )$.
  Key ideas in our method can be applicable to another method which has a suitable mathematical structure for our ideas.

 \section*{Acknowledgement} 
 
  The authors would like to thank to Prof. Yoshinobu Kuramashi, Associate Prof. Shinji Takeda, Project Associate Prof. Tsuyoshi Okubo, Dr. Satoshi Morita, Dr. Yoshifumi Nakamura, Dr. Yusuke Yoshimura, Dr. Yasunori Futamura and Mr. Shinichiro Akiyama
  for useful suggestions and meaningful discussion.
  Numerical experiments in this work are performed using Oakforest-PACS system in Joint Center for Advanced High Performance Computing.
  This research used computational resources of the K computer through the HPCI System Research Project (Project ID: hp180225)
                                  and the Fujitsu PRIMERGY CX600M1/CX1640M1 (Oakforest-PACS) in the Information Technology Center, The University of Tokyo.
  This work was supported by JSPS KAKENHI Grant Number JP18H03250.

\end{document}